\documentclass[12pt]{article}
\pdfoutput=1

\usepackage[T1]{fontenc}
\usepackage{verbatim}
\usepackage{float}
\usepackage{amsthm}
\usepackage{amsmath}
\usepackage{amssymb}
\usepackage{graphicx}
\usepackage{color}
\usepackage{url}
\usepackage{caption}
\usepackage{subcaption}
\usepackage{mathtools} 
\usepackage{stackrel}

\newcommand{\E}{\mathbb{E}}

\newcommand{\ep}{\varepsilon}

\newcommand{\1}{\mathbf{1}}

\newcommand{\Poisson}{\text{Poisson}}

\newcommand{\revised}[1]{{\color{black}{#1}}}
\newcommand{\R}{\mathbb{R}}

\makeatletter

\newcommand{\RL}{\mathbb{R}^L}

\newcommand{\RN}{\mathbb{R}^N}

\newcommand{\one}{\mathbf{1}} 
\newcommand{\be}
{\begin{equation}}
\newcommand{\ee}
{\end{equation}}
\newcommand{\aseq}{\stackrel{a.s.}{=}}

\theoremstyle{plain}
\newtheorem{thm}{\protect\theoremname}[section]
\theoremstyle{definition}

\theoremstyle{remark}

\theoremstyle{plain}
\newtheorem{lem}[thm]{\protect\lemmaname}
\newtheorem*{lem*}{Lemma}
\theoremstyle{remark}

\theoremstyle{plain}
\newtheorem{corollary}[thm]{\protect\corollaryname}
\theoremstyle{plain}

\theoremstyle{plain}
\newtheorem{proposition}[thm]{\protect\propositionname}
\providecommand{\claimname}{Claim}
\providecommand{\definitionname}{Definition}
\providecommand{\lemmaname}{Lemma}
\providecommand{\remarkname}{Remark}
\providecommand{\theoremname}{Theorem}
\providecommand{\corollaryname}{Corollary}
\providecommand{\propositionname}{Proposition}
\providecommand{\conjecturename}{Conjecture}

\usepackage{authblk}

\usepackage[margin=3cm]{geometry}

\allowdisplaybreaks
\numberwithin{equation}{section}

\RequirePackage[colorlinks,citecolor=blue,urlcolor=blue,linkcolor=blue]{hyperref}

\begin{document}


\title{Multi-target detection with application to cryo-electron microscopy}

\author[a]{Tamir Bendory}
\author[b]{Nicolas Boumal} 
\author[c]{William Leeb}
\author[a,b]{Eitan Levin}
\author[a,b]{Amit Singer}

\affil[a]{The Program in Applied and Computational Mathematics, Princeton University, Princeton, NJ, USA}
\affil[b]{Department of Mathematics, Princeton University, Princeton, NJ, USA}
\affil[c]{School of Mathematics, University of
	Minnesota, Minneapolis, MN, USA }

\maketitle

\begin{abstract}

We consider the multi-target detection problem of recovering a set of signals that appear 
multiple times at unknown locations in a noisy measurement.
In the low noise regime, one can estimate the signals by first detecting occurrences, then clustering and averaging them.
In the high noise regime, however, neither detection nor clustering can be performed reliably, so that strategies along these lines are destined to fail.  
Notwithstanding, using autocorrelation analysis, we show that the impossibility to detect and cluster signal occurrences in the presence of high noise does not necessarily preclude signal estimation.
Specifically, to estimate the signals, we derive simple relations between the autocorrelations of the observation and those of the signals. These autocorrelations can be estimated accurately at any noise level given a sufficiently long measurement. 
To recover the signals from the observed autocorrelations, we  solve a set of polynomial equations through nonlinear least-squares. 
We provide analysis regarding well-posedness of the task, and demonstrate numerically the effectiveness of the method in a variety of settings.

The main goal of this work is to provide theoretical and numerical support for a recently proposed framework to image 3-D structures of biological macromolecules using cryo-electron microscopy in extreme noise levels. 
 
\end{abstract}

\section{Introduction} \label{sec:intro}

We consider the \emph{multi-target detection} problem of recovering a set of $K$ signals that appear 
multiple times at unknown locations in a noisy measurement.
Let ${x_1,\ldots,x_K\in\RL}$ be the sought signals and let $y\in\RN$ be the observed data, where we assume $N$ is  much larger than $L$. 
Let  $s[i]$ count the number of signal occurrences whose first entry is positioned at $y[i]$. Each of those $s[i]$ signals is chosen among $x_1, \ldots, x_K$ according to some (possibly unknown) distribution over $\{1,\ldots,K\}$. 
If signal occurrences overlap, they interfere additively. 
With additive white Gaussian noise, the measurement model can be written as 
\begin{align} 
	y  &=  \sum_{k=1}^K s_k \ast x_k + \varepsilon, &\varepsilon   \sim \mathcal{N}(0,\sigma^2 I_N),
	\label{eq:model}
\end{align}
where $\ast$ denotes linear convolution, and $s_k[i]$ indicates the number of occurrences of $x_k$ starting at $y[i]$, so that $s =  s_1+\cdots+s_K$. Explicitly, with zero-based indexing, 
\begin{align*}
	y[i] & = \sum_{k=1}^{K} \sum_{j = 0}^{L-1} s_k[i-j] x_k[j] + \varepsilon[i].
\end{align*}
%
The goal is  to estimate $x_1,\ldots,x_K$ from $y$. 
In parts of the paper, we focus on the case $K = 1$, called the \emph{homogeneous} case; the case $K \geq 2$ is called \emph{heterogeneous}.
This idealized setup appears in several scientific applications, including structural biology~\cite{bendory2018toward} (as we detail below), spike sorting~\cite{lewicki1998review}, passive radar~\cite{gogineni2017passive}, and system identification~\cite{ljung1998system}. 

In the low noise regime (small $\sigma$), a valid strategy is to first detect the signal occurrences in $y$ (that is, estimate $s$), cluster them (that is, separate $s$ into $s_1,\ldots,s_K$), and solve  standard deconvolution problems.
Crucially, we focus on the high noise regime, where \emph{reliable detection of signal occurrences is  impossible}~\cite{bendory2018toward,aguerrebere2016fundamental}.
This limitation does not, however, preclude estimation of the signals $x_1,\ldots,x_K$, as we show in this paper. In this setting, we refer to  $s_1,\ldots,s_K$  as \emph{nuisance variables}: knowing them would certainly help, but we do not aim to estimate them.

In order to recover the signals in the high noise regime, we use autocorrelation analysis.
At any noise level, the autocorrelations of the observation can be estimated to any desired accuracy for sufficiently large  $N$. 
This computation is straightforward and requires only one pass over the data.
The underlying principle is to relate the autocorrelations of the observation $y$ to the autocorrelations of $x_1,\ldots,x_K$.

Below we describe two generative models for $s$.  
In these models, the relationship between the autocorrelations of $y$ and those of $x_1,\ldots,x_K$ depends on $s_1,\ldots,s_K$ only through their expected sums, that is, the expected total number of occurrences of each signal.
To estimate the signals and occurrence counts from the computed autocorrelations, we solve a nonlinear least-squares problem as explained in Section~\ref{sec:numerics}. 

The multi-target detection problem  is an instance of  
\emph{blind deconvolution}---a longstanding problem arising in a variety of engineering and scientific applications, such as astronomy, communication, image deblurring, system identification and optics; see~\cite{jefferies1993restoration,shalvi1990new,ayers1988iterative,abed1997blind}, to name a few. 
Different variants of the blind deconvolution problem have been thoroughly analyzed recently~\cite{ahmed2014blind,li2016identifiability,li2016rapid,lee2017blind,ling2017blind,kuo2019geometry}. In clear contrast to multi-target detection, these works focus on the low noise regime and aim to estimate both unknown signals (in our setting, this means estimating both $x_k$'s and $s_k$'s).

\subsection*{Models for target distribution}

We consider two  models for the distribution of signal occurrences in the observation, that is, for $s_1, \ldots, s_K$.

\paragraph{The well-separated model.}

As a first setup, we allow any generative model for $s$ which meets the following separation requirement: $s$ is binary, and
\begin{equation}
\textrm{If } s[i] = 1 \textrm{ and } s[j] = 1 \textrm{ for } i \neq j, \textrm{ then } |i - j| \geq 2L-1.
\label{eq:spacing}
\end{equation}
In words: the starting positions of any two occurrences  must be separated by at least $2L-1$ positions, so that their end points are necessarily separated by at least $L-1$ signal-free (but still noisy) entries in the data.
Furthermore, we require that the last signal occurrence in $y$ is also followed by at least $L-1$ signal-free entries.
This property ensures that correlating $y$ with versions of itself shifted by at most $L-1$ entries does not involve correlating distinct signal occurrences. Once $s$ is determined, for each position $i$ such that $s[i] = 1$, one of the signals $x_k$ is selected independently at random, and accordingly we set $s_k[i] = 1$. As a result, the only properties of $s_1, \ldots, s_K$ that affect the autocorrelations of $y$ (for shifts up to $L-1$) are the total number of occurrences of the distinct signals: their individual and relative locations do not intervene. We detail this in Section~\ref{sec:AC_analysis}.


\paragraph{The Poisson model.}

If the separation condition is violated, more knowledge about the location distribution is necessary to disentangle the autocorrelations of $y$. To that effect, we analyze a Poisson generative model.

Specifically, for each position $i$, the number $s[i]$ of signal occurrences starting at that position is drawn independently from a Poisson distribution with parameter $\gamma/L$, that is, ${s[i] \overset{i.i.d.}{\sim} \Poisson(\gamma / L)}$ for some parameter $\gamma > 0$. Then, $s[i]$ is split into $s_1[i] + \cdots + s_K[i]$ by selecting $s[i]$ signals among $x_1, \ldots, x_K$, independently at random  following a fixed distribution over $\{1,\ldots,K\}$.
It is possible for more than one occurrence of the same $x_k$ to start at position~$i$.
As for the well-separated model, the autocorrelations of $y$ under this model depend weakly on $s$ and $s_1, \ldots, s_K$, essentially through the parameter $\gamma$: see Section~\ref{sec:AC_analysis}.

\subsection*{Extensions} \label{sec:extensions}

Extending the problem setup and autocorrelation analysis to signals in more than one dimension is straightforward: see the discussion in Section~\ref{sec:high_dimensions} and numerical experiments in Section~\ref{sec:numerics}.

Likewise, it is easy to extend the model to situations where the signal occurrences are sampled from a general distribution rather than from a finite set of choices $x_1, \ldots, x_K$. The finite setup corresponds to the following distribution for signal occurrences:
\begin{align}
	x \sim \sum_{k=1}^K \pi_k \delta(x - x_k),
	\label{eq:heterogeneousdistribution}
\end{align}
where $\delta(x - x_k)$ is a Dirac delta (a point mass) located at $x_k$, and $(\pi_1, \ldots, \pi_K)$ encodes the discrete distribution over $\{1, \ldots, K\}$.
In the generalized setup, the goal is to estimate the distribution (possibly defined by a finite set of parameters). In particular, this allows for continuous distributions of targets. We adopt this perspective when deriving the autocorrelations in Section~\ref{sec:AC_analysis}.

In the next section, we show how this flexibility allows us to model an important imaging problem in structural biology.

\section{Connection with single-particle reconstruction via cryo-electron microscopy} \label{sec:cryoem}

Cryo-electron microscopy (cryo-EM)  has recently joined X-ray crystallography and nuclear magnetic resonance (NMR) spectroscopy as a high-resolution structural method for biological macromolecules~\cite{frank2006three,kuhlbrandt2014resolution,bartesaghi20152}.
In a cryo--EM experiment, biological samples (e.g., macromolecules, viruses) are rapidly frozen in a thin layer of vitreous ice.
The microscope produces 2-D tomographic images of the samples embedded in the ice, called \emph{micrographs}. Each micrograph contains multiple tomographic projections of the samples at unknown locations and under unknown viewing directions. 
Importantly, the electron dose must be kept low to mitigate  radiation damage, inducing high noise levels.
The goal is to reconstruct 3-D models of the molecular structures from the micrographs. 
Since  cryo-EM produces images of individual particles, it can elucidate multiple  structures simultaneously.
This is in clear contrast to  X-ray and NMR, which aggregate information from an ensembles of
particles.

Considering the extensions described in the previous section, we can phrase a simplified generative model for micrographs (the observation $y$) within our framework. Specifically, locations are chosen in the 2-D plane of the image, corresponding to $s$: this is where the molecules are fixed in the plane of the ice layer. At each selected location, a 2-D tomographic projection of a molecule (a signal occurrence) is added in the observation $y$. This signal is drawn from a probability distribution described by a discrete number of parameters which correspond to the sought 3-D structure, as follows.

The 3-D structure $V$ (the target parameter) can be expanded into a linear combination of basis functions (for example, spherical harmonics for the spherical part and Bessel functions for the radial part): the coefficients of this expansion are the unknowns. Then, a rotation $R_\omega$ is applied to the volume (to model viewing directions) according to a (possibly unknown) distribution of $\omega$ over $SO(3)$ (the group of 3-D rotations). With tomographic projection denoted by $P$, $x$ is a random signal, related to the distribution of $\omega$ through: 
\begin{align}
	x & = P(R_\omega V).
	\label{eq:cryoEM}
\end{align}
By further allowing $V$ itself to be a random signal as well, this model can also encode a mixture of structures in the biological sample. This mixture might also be continuous, corresponding to continuous conformational variability.

All contemporary methods for single particle reconstruction using cryo-EM  split the reconstruction procedure into two main  stages.
The first stage, called \emph{particle picking},  detects and extracts the particle projections from the micrographs. Given the projections, the second stage aims to reconstruct a 3-D model of the molecular structure, usually using an expectation-maximization algorithm~\cite{scheres2012relion}. 
Crucially, reliable detection of individual particles is impossible in a highly noisy environment. This fact has been recognized early on by the cryo-EM community. 
Particularly, in~\cite{henderson1995limitations,glaeser1999electron}, it was reasoned that particle picking is impossible for molecules below a certain weight (below~$\sim$50 kDa). 

Even if particle picking is feasible, procedures may be affected by \emph{model bias}.
Particle picking algorithms are typically based on correlating the micrograph with several templates. Areas highly correlated with some of the templates are assumed to contain projections: these yield the picked particles. 
Clearly, the picked particles depend heavily on the chosen templates: different templates may lead to different picked particles, hence, which is more problematic, to different 3-D structure reconstructions.
The cryo-EM community is well aware of this potential pitfall~\cite{vanheel1992correlation,henderson2013avoiding,vanheel2013finding}, which was notably exemplified \revised{by} the ``Einstein from noise'' experiment~\cite{shatsky2009method}.

A recent work of the authors suggests a methodology to bypass particle picking and  reconstruct the 3-D structure directly from the micrograph~\cite{bendory2018toward}.
Based on autocorrelation analysis, it was shown that---at least in principle---the limits high noise regimes 
impose on particle picking  do not necessarily translate into limits on 3-D reconstruction.
The main goal of  the present paper is to provide theoretical and numerical support for this approach, in a simplified setting.

In the next section, we introduce autocorrelation analysis in detail, focusing on the multi-target detection model. 
We mention that similar ideas in  cryo-EM can be traced back to a seminal paper of Zvi Kam~\cite{kam1980reconstruction}. Kam proposed autocorrelation analysis for \mbox{3-D} reconstruction, under the assumption of picked, perfectly centered, particles. Kam's method has been extended and used in X-ray free electron lasers (XFEL) and cryo-EM~\cite{liu2013three,kurta2017correlations,levin20173d,von2018structure}.  
In order to investigate the computational and statistical properties of Kam's method, a series of papers have studied a simplified model, 
called   \emph{multi-reference alignment}~\cite{bandeira2014multireference,bendory2017bispectrum,bandeira2017optimal,perry2017sample,bandeira2017estimation,abbe2017multireference}.
We follow the same line of research by considering the multi-target detection as an abstraction to the application of reconstructing 3-D structures directly from the micrograph~\cite{bendory2018toward}.

\section{Autocorrelation analysis} \label{sec:AC_analysis}

In what follows, we consider autocorrelations of both the observation $y$ and of the signal occurrences in $y$. As per our discussion of extensions, the signal occurrences may be sampled from a discrete set $\{x_1, \ldots, x_K\}$ (as in~\eqref{eq:heterogeneousdistribution}), or from a more general distribution. Accordingly, we define autocorrelations broadly for a random signal $z$ of length $M$. For our purposes, this will be applied both to signal occurrences (of length $L$) and to $y$ (of length $N$).

For a random signal $z\in \R^M$, the autocorrelation of order $q = 1, 2, \ldots$ is given for any integer shifts $\ell_1, \ldots, \ell_{q-1}$ by
\begin{equation}
a_z^q[\ell_1,\ldots,\ell_{q-1}]   = \E_z\left\{\frac{1}{M} \sum_{i=-\infty}^{\infty} z[i]z[i+\ell_1]\cdots z[i+\ell_{q-1}]\right\},
\label{eq:ac_general}
\end{equation}
where the expectation is taken with respect to the distribution of $z$. Indexing out of bounds is zero-padded, that is, $z[i] = 0$ for $i$ out of the range $0, \ldots, M-1$.
Explicitly, the first-, second- and third-order autocorrelations are given by: 
\begin{align}  
a_z^1 & = \E_z\left\{\frac{1}{M} \sum_{i=0}^{M-1} 
z[i]\right\}, \nonumber\\
a_z^2[\ell] & = \E_z\left\{\frac{1}{M} \sum_{i = \max\{0, -\ell\}}^{M-1 + \min\{0, -\ell\}} z[i]z[i+\ell]\right\}, \label{eq:ac_special} \\
a_z^3[\ell_1,\ell_2] & = \E_z\left\{\frac{1}{M} \sum_{i = \max\{0, -\ell_1, -\ell_2\}}^{M-1 + \min\{0, -\ell_1, -\ell_2\}} z[i]z[i+\ell_1]z[i+\ell_2]\right\}. \nonumber
\end{align}
Since  autocorrelations depend only on the differences between indices, they obey the following symmetries: $$a_z^2[\ell] = a_z^2[-\ell],$$ and
$$a_z^3[\ell_1,\ell_2] = a_z^3[\ell_2,\ell_1]=a_z^3[-\ell_1,\ell_2-\ell_1].
$$
In particular, for
$x$ sampled from $\{ x_1, \ldots, x_K \}$ with probabilities $(\pi_1, \ldots, \pi_K)$ as in~\eqref{eq:heterogeneousdistribution}, the autocorrelations of $x$ are given in explicit form in terms of those of the deterministic signals $x_1, \ldots, x_K$ as:
\begin{align}
	a_x^q  & = \sum_{k=1}^{K} \pi_ka_{x_k}^q.
	\label{eq:mixedautocorr}
\end{align}
Explicit expressions of the autocorrelations for the more involved model of  cryo-EM~\eqref{eq:cryoEM} are given in~\cite{bendory2018toward}.

We are given one observation (one realization) of $y$. Thus, we cannot compute the autocorrelations of $y$ exactly as they involve taking an expectation against the distribution of $y$. However, by the law of large numbers, as $N$ grows to infinity  the empirical autocorrelations of $y$ almost surely (a.s.) converge to the actual (population) autocorrelations of $y\in \RN$, that is, 
\begin{align}
\lim_{N \to \infty} \frac{1}{N} \sum_{i=-\infty}^{\infty} y[i]y[i+\ell_1]\cdots y[i+\ell_{q-1}] \aseq a_y^q[\ell_1, \ldots, \ell_{q-1}].
\end{align}
This provides a concrete means of estimating the quantities $a_y^ q$. In the remainder of this section, we relate the observables $a_y^q$  to the unknowns $a_x^q$, first under the well-separated model, then under the Poisson model.

\subsection{Autocorrelations under the well-separated model}

Under the separation condition~\eqref{eq:spacing}, the relation between autocorrelations of the observation $y$ and those of $x$ is particularly simple, as we now show. It is useful to introduce some notation: let $\vert s\vert = \sum_i s[i]$ denote the number of signal occurrences in $y$, and let
\begin{equation}
\gamma  = \frac{|s| L}{N}.
\label{eq:gamma}
\end{equation}
This $\gamma$ is the fraction of entries of $y$ occupied by signal occurrences. 
The separation condition  imposes $\gamma\leq\frac{L}{2L-1}\approx 1/2$. 
For the heterogeneous model~\eqref{eq:model},  one can define
\begin{equation}
\gamma_k  = \frac{|s_k| L}{N},
\label{eq:gamma_k}
\end{equation}
where $\vert s_k\vert = \sum_i s_k[i]$ so that $\gamma=\sum_k \gamma_k$.

Owing to the separation condition, when correlating $y$ with shifted versions of itself for shifts in $0, \ldots, L-1$, any given occurrence of $x$ in $y$ is only ever correlated with itself, and never with another occurrence. As a result, the autocorrelations of $y$ depend on the corresponding autocorrelations of $x$, the noise level $\sigma$ and the density $\gamma$ (which is a weak dependence on the support signal $s$). Specifically, we show the following identities in Appendix~\ref{sec:autocorrelation_computation}:
\begin{align} 
	a_y^1 & = \gamma a_{x}^1, \label{eq:mean_micrograph} \\
	a_y^2[\ell] & = \gamma a_{x}^2[\ell] + \sigma^2\delta[\ell], \label{eq:ac2_micrograph}\\
	a_y^3[\ell_1,\ell_2] & = \gamma a_{x}^3[\ell_1,\ell_2]  + \sigma^2\gamma a_{x}^1  \big(\delta[\ell_1]+\delta[\ell_2]
	+\delta[\ell_1-\ell_2]\big), \label{eq:ac3_micrograph}
\end{align}
where $\delta[0] = 1$ and $\delta[\ell \neq 0] = 0$, and indices $\ell, \ell_1, \ell_2$ are in the range $0 \leq \ell \leq L-1$. Terms proportional to $\sigma^2$ are due to noise. If $\sigma$ is known, they can be handled easily. If $\sigma$ is unknown, one can either estimate it form the data, or one can ignore the few entries of the autocorrelations that are affected by $\sigma$---one in $a_y^2$ and $3L-2$ in $a_y^3$, a relatively small number in both cases.

We show in Section~\ref{sec:theory_homogeneous_well_separated} 
that $x,\gamma$ and $\sigma$ can be identified uniquely from the observed autocorrelations for the homogeneous case $K = 1$ (all signal occurrences are the same). 

\subsection{Autocorrelations under the Poisson model} \label{sec:ac_poisson}

In this section, we give the expressions for the autocorrelations of the observed signal $y$ under the Poisson model. 
 We first note that the expected total number of signals occurring in $y$ is equal to $(N - L + 1)\gamma / L$, and consequently the total lengths of all signals (including overlaps) divided by $N$ is equal to
\begin{align}
\frac{N - L + 1}{N} \cdot \frac{\gamma  }{L}  \cdot L 
\quad \overset{N\to\infty}{\longrightarrow} \quad \gamma.
\end{align}
Therefore, in the large $N$ limit, similarly to the role of $\gamma$ in the well-separated model~\eqref{eq:gamma}, the parameter $\gamma$ may be interpreted as the signal density.

In Appendix~\ref{sec:proof_prop_poisson}, we prove the following expressions for the autocorrelations of $y$ under the Poisson model. As in the well-separated model, the observed autocorrelations do not depend on individual occurrences of the signal, but only on the distribution of $x$ itself, and the parameters $\gamma$ and $\sigma$: 
\begin{align}
	a_y^1  & =  \gamma a_{x}^1, \label{eq:mean_micrograph2} \\
	a_y^2[\ell] & = \gamma a_x^2[\ell] + \sigma^2 \delta[\ell] + (\gamma a_x^1)^2, \label{eq:ac2_micrograph2} \\
	a_y^3[\ell_1,\ell_2]  & =  \gamma a_x^3[\ell_1,\ell_2] + \sigma^2 \gamma a_x^1 \big(\delta[\ell_1]+\delta[\ell_2] +\delta[\ell_1-\ell_2] \big) \nonumber \\
						  & \qquad + (\gamma a_x^1)^3 + \gamma a_x^1 \cdot ( \gamma a_x^2[\ell_1] + \gamma a_x^2[\ell_2] + \gamma a_x^2[\ell_2-\ell_1]). \label{eq:ac3_micrograph2}
\end{align}
Note that from those autocorrelations, it is easy to retrieve the autocorrelations of the well-separated model for all entries unaffected by $\sigma$. If $\sigma$ is known, then it is true for all entries. 

In the homogeneous case ($K=1$), we show in Section~\ref{sec:theory_homogeneous_poisson} that $a_y^1,a_y^2$ and $a_y^3$ identify uniquely the signal $x$, the Poisson parameter $\gamma$, and the noise level $\sigma$ for generic $x$. 

\section{Theory}

We begin this section by showing that, in the homogeneous case ($K=1$), under both the well-separated and the Poisson models, the first three observed autocorrelations  identify the (deterministic) signal $x$, the density parameter $\gamma$, and the noise level $\sigma^2$.  Then, for the heterogeneous case ($K \geq 2$), we bound from above the number $K$ of signals that can be recovered from those autocorrelations as a function of the signal length $L$. Finally, we briefly discuss multi-dimensional signals.

Before that, we start by showing that in the homogeneous case a deterministic signal $z\in\RL$ is identified uniquely by its second and the third autocorrelations.
 Indeed, assuming $z[0]$ and $z[L-1]$ are nonzero, we can recover $z$ explicitly by 
\begin{equation}
z[k]  = \frac{z[0]z[k]z[L-1]}{z[0]z[L-1]} = \frac{a_z^3[k,L-1]}{a_z^2[L-1]},
\label{eq-uniqueness}
\end{equation}
for $k = 0, \ldots, L-1$. If $z[0]$ or $z[L-1]$ are equal to zero, then $a_z^2[L-1] = 0$ and we can use that indication to shrink $L$.
This proves the following useful fact:
\begin{proposition} \label{prop:uniqueness}
	A deterministic signal $z\in\RL$ is determined uniquely by  $a_z^2$ and $a_z^3$. 
\end{proposition}

Note that the procedure described in~\eqref{eq-uniqueness} is not numerically stable: if $z[0]$ or $z[L-1]$ are close to 0, recovery of $z$ is sensitive to errors in the autocorrelations. In practice, we recover $z$ by fitting it to its autocorrelations using a nonconvex least-squares procedure, which is empirically robust to additive noise. In prior work, we have observed similar phenomena for the related problem of multi-reference alignment~\cite{bendory2017bispectrum,boumal2017heterogeneous,abbe2017multireference}.

\subsection{Guarantees for the homogeneous well-separated model} \label{sec:theory_homogeneous_well_separated}
 
The observed moments $a_y^1,a_y^2$ and $a_y^3$ under the well-separated model do not immediately yield the autocorrelations  of the signal $x$; rather, the two are related by the noise level $\sigma$ and the signal density $\gamma$. We will show, however, that all parameters---$x, \gamma$ and $\sigma^2$---are still identified uniquely by the observed moments of $y$.

First, we observe that if  the noise level $\sigma$ is known, generally, one can estimate $\gamma$ from the first two moments of the micrograph.
The other direction is true as well: if $\gamma>0$ is known, then one can estimate $\sigma^2$.
 The proof is provided in Appendix~\ref{sec:proof_prop_gamma}.

\begin{proposition} \label{prop:gamma}
	Assume the separation condition~\eqref{eq:spacing} holds and $K=1$ (all signal occurrences in $y$ are identical, up to noise). If the mean of $x$ is nonzero, then  
	\begin{equation}
	\gamma  = \frac{L (a^1_y)^2}{a_y^2[0] + 2\sum_{\ell = 1}^{L-1}a_y^2[\ell]-\sigma^2},
	\end{equation}
	meaning $\gamma$ can be determined from $\sigma$ (and vice versa) using the observables $a_y^1, a_y^2$.
\end{proposition}

Using third-order autocorrelation information of $y$, both the ratio $\gamma$ and the noise $\sigma$ can be determined simultaneously. For the following results, when we say that a result holds for a ``generic'' signal $x$, we mean that the set of signals which cannot be determined by these measurements
has Lebesgue measure zero. 
In particular, this means that we can recover
almost all signals with the given measurements. The proof is provided in Appendix~\ref{sec:proof_prop_gamma_sigma}.

\begin{proposition} \label{prop:gamma_sigma}
	Assume $L \geq 3$, $K=1$ and assume that the separation condition~\eqref{eq:spacing} holds. 
	Then, the observed autocorrelations $a_y^1,a_y^2$ and  $a_y^3$ determine the ratio $\gamma$ and noise level $\sigma$ uniquely for a generic signal $x$. If $\gamma > \frac{1}{4}$, then this holds for any signal $x$ with nonzero mean. 
\end{proposition}

From Propositions~\ref{prop:uniqueness} and~\ref{prop:gamma_sigma} we   deduce the following:
\begin{corollary}
	Assume $L \geq 3$ and $K=1$.
	Under the separation condition~\eqref{eq:spacing}, the signal $x$, the ratio $\gamma$, and the noise level $\sigma$ are determined from the first three autocorrelation functions of $y$ if either the signal $x$ is generic, or $x$ has nonzero mean  and $\gamma > \frac{1}{4}$.
\end{corollary}
%

\subsection{Guarantees for the homogeneous Poisson model} \label{sec:theory_homogeneous_poisson}

Similarly to the homogeneous well-separated model, the observed autocorrelations under the Poisson model identify $x,\gamma$ and $\sigma^2$ uniquely.
Opposed to Proposition~\ref{prop:gamma_sigma}, these quantities can be computed explicitly. 
In Appendix~\ref{sec:proof_uniqueness_poisson} we prove the following result:

\begin{proposition} \label{prop:uniqueness_poisson}
Under the homogeneous Poisson model, the signal $x$, the noise level $\sigma^2$ and the Poisson parameter $\gamma$ are identified uniquely \revised{from the observed autocorrelations}. 
In particular,
	\begin{align}
\gamma &= L \frac{(\gamma a_y^1) (a_y^2[1] - (a_y^1)^2)}
{\sum_{\ell=0}^{L-1} \gamma a_x^3[1,\ell] 
	+ \sum_{\ell=2}^{L-1} \gamma a_x^3[\ell,\ell+1]}, \\
\sigma^2 &= a_y^2[0] + 2\sum_{\ell = 1}^{L-1}a_y^2[\ell]-\frac{L (a^1_y)^2}{\gamma}
- (2 L - 1) (a_y^1)^2,
\end{align}
where \revised{$\gamma a_x^3$ is indeed observable since}
\begin{align}
\gamma a_x^3[\ell_1,\ell_2]
&= a_y^3[\ell_1,\ell_2] - (a_y^1)^3
- a_y^1 \cdot \big(a_y^2[\ell_1]  
+ a_y^2[\ell_2] + a_y^2[\ell_1-\ell_2]- 3(a_y^1)^2 \big).
\end{align}

\end{proposition}

\subsection{Elementary limitations of the heterogeneous case}
\label{sec:heterogeneity}

In the heterogeneous model~\eqref{eq:model}, the unknowns are $K$ signals of length $L$, together with their densities $\gamma_1, \ldots, \gamma_K$~\eqref{eq:gamma_k} (equivalently: the distribution $\pi$ and overall density $\gamma$) and possibly the noise level $\sigma$. To estimate these parameters, we must collect at least as many independent equations. Within our framework, polynomial equations are provided by the observable autocorrelations, which correspond to mixed autocorrelations of the unknowns as per~\eqref{eq:mixedautocorr}. In this section, following~\cite{boumal2017heterogeneous}, we count how many equations the first three autocorrelations may provide in the best case (discounting symmetries). This leads to a straightforward information-theoretic upper bound on the number $K$ of signals which can be estimated, as a function of $L$. This is only an upper bound, though a bound of the same type was shown to be tight in a similar setting~\cite{bandeira2017estimation}.
The counting is based on the autocorrelations of the well-separated model. For entries independent of $\sigma$, the autocorrelations of the Poisson process contain the same information as those of the well-separated model. If $\sigma$ is known, this holds true for all entries; see Section~\ref{sec:ac_poisson}.

The first-order autocorrelation $a_y^1$~\eqref{eq:mean_micrograph} provides one equation. For second-order autocorrelations $a_y^2[\ell]$~\eqref{eq:ac2_micrograph}, if $\sigma$ is known we obtain $L$ equations with $\ell$ ranging from 0 to $L-1$. If $\sigma$ is unknown, we may disregard $a_y^2[0]$ (the only entry affected by $\sigma$) and still collect $L-1$ equations. Similarly, for third-order autocorrelations, $a_y^3[\ell_1, \ell_2]$~\eqref{eq:ac3_micrograph} with $0 \leq \ell_1, \ell_2 \leq L-1$ such that $\ell_2 \leq \ell_1$ includes all relevant entries for our purpose (this accounts for symmetries), providing $\frac{(L+1)(L+2)}{2}-2$ equations in total. 
If we further exclude any entries such that $\ell_1, \ell_2$ or $\ell_1 - \ell_2$ are zero to avoid the need to estimate $\sigma$, there are $\frac{(L-1)(L-2)}{2}$ remaining entries.

Hence, if $\sigma$ is known we collect
\begin{align*}
	1 + L + \frac{(L+1)(L+2)}{2}-2 = \frac{1}{2} L (L+5)
\end{align*}
equations, while if it is unknown and we choose not to estimate it, then we collect
\begin{align*}
	1 + (L-1) + \frac{(L-1)(L-2)}{2} = \frac{1}{2} L (L-1) + 1
\end{align*}
equations in total. Of course, there may be redundancy in these equations: we aim only to provide an upper bound. 

Since we aim to estimate $KL$ parameters for the $K$ signals of length $L$, plus $K$ parameters for the densities $\gamma_k$, 
 there are $K(L+1)$ unknowns. As a result, an absolute upper bound on $K$ such that the estimation problem may be solvable is
\begin{align*}
	K \leq \frac{L(L+5)}{2(L+1)}
\end{align*}
for the case of $\sigma$ known, and
\begin{align*}
	K \leq \frac{ L (L-1) + 1}{2(L+1)}
\end{align*}
for the case of $\sigma$ unknown and not estimated. Overall, this indicates that, at best, approximately $L/2$ signals and their densities can be recovered from the first three mixed autocorrelations. Based on related results in~\cite{bandeira2017estimation}, we expect that as many as $L/2$ signals can indeed be estimated, though possibly not with computationally tractable estimators.

\subsection{Autocorrelations in higher dimensions} \label{sec:high_dimensions}

Autocorrelations in $d$ dimensions are defined for  $\ell_1,\ldots,\ell_{q-1}\in\mathbb{Z}^d$  as 
\begin{equation}
a_z^q[\ell_1,\ldots,\ell_{q-1}]   = \E\left\{\frac{1}{L^d} \sum_{i\in\mathbb{Z}^d} z[i]z[i+\ell_1]\cdots z[i+\ell_{q-1}]\right\}.
\label{eq:ac_d_dimension}
\end{equation}
Interestingly, for the homogeneous case ($K=1$) in dimensions greater than one, almost all  signals are determined uniquely from their second-order autocorrelation, up to two symmetries: sign (or phase for complex signals) and reflection through the origin (with conjugation in the complex case)~\cite{hayes1982reconstruction}. 
If the mean of the signal is available and non-zero, the sign symmetry can be resolved. However, determining the reflection symmetry still requires additional information, beyond the second-order autocorrelation.
The case of 1-D signals is fundamentally different: generally there are $2^{L-2}$ signals with the same second-order autocorrelation (after eliminating  symmetries)~\cite{beinert2015ambiguities,bendory2017fourier}.

This uniqueness result for two and three-dimensional signals is the basis of a popular imaging technique called coherent diffraction imaging (CDI). In CDI, an object is
illuminated with a coherent wave, and the far field diffraction pattern  is measured, corresponding to the object's Fourier magnitude~\cite{miao1999extending,shechtman2015phase}. 
If the diffraction pattern is over-sampled by at least  twice the Nyquist frequency,  the data is equivalent to the signal's second-order autocorrelation. In this context, the computational problem of recovering the signal from its second-order autocorrelation is usually referred to as \emph{phase retrieval} or the \emph{phase problem}.
\revised{However, for 2-D images it has been shown recently that the problem is  ill-conditioned unless the support of the image is known exactly~\cite{barnett2018geometry}}. That is, there exist other images whose second-order autocorrelations agree up to machine precision.

\section{Algorithms and numerical experiments} \label{sec:numerics}

In this section, we present three numerical experiments. In the first two experiments, we consider the heterogeneous model~\eqref{eq:model}. First, with a fixed noise level, we show how the estimation quality improves as the length of the observation grows. Second, we explore how many signals can be recovered as a function of their length, in an infinite data regime where effects of the noise have been averaged out. In the last experiment, we extend our model to 2-D signals.
We run the experiments on a shared computer with 144 logical CPUs of type Intel(R) Xeon(R)
CPU E7-8880 v3 $@$ 2.30GHz and 755Gb of RAM; we use at most 72 of these CPUs.
The code for all experiments is available at \url{https://github.com/PrincetonUniversity/BreakingDetectionLimit}. 

\revised{In this section, to assess the quality of our reconstruction, we compare against the ground truth. In a realistic setting, the ground truth is of course not available, which raises the question of how one can validate the results. A common technique used in cryo-EM is to split the data in two halves, produce two independent reconstructions based on these halves, then to compare the two reconstructions. It is clear how the same technique can be applied to our setting.}

\subsection{Experiment 1} 
\label{sec:XP1}

\begin{figure}[t]
	\centering
	\includegraphics[width=1\linewidth]{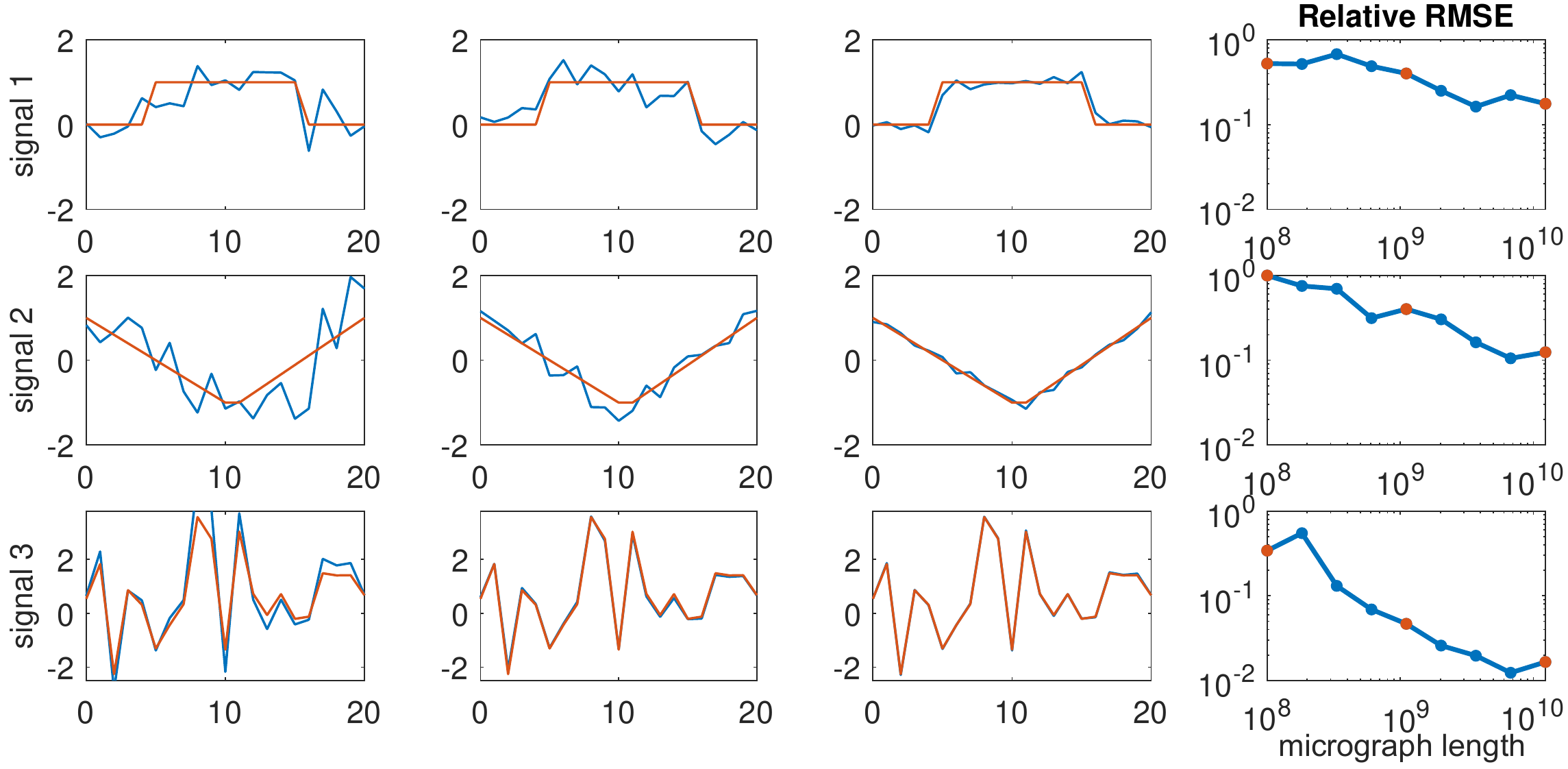}
	\caption{Experiment described in Section~\ref{sec:XP1}. For a fixed noise level $\sigma = 3$ and a fixed set of $K = 3$ signals of length $L = 21$, an observation $y$ of length $N = 1.23 \cdot 10^{10}$ is generated according to~\eqref{eq:model} in the well-separated model, with fixed  occurrence probabilities. Each row corresponds to one of the signals. The last column shows evolution of the relative root mean squared error in estimating each signal, as a longer and longer subset of $y$ is observed. Red dots mark the three snapshots that are illustrated in columns 1--3: red signals are the ground truth and blue signals are the estimators.}
	\label{fig:1Dheterosignals}
\end{figure}

For the experiment depicted in Figure~\ref{fig:1Dheterosignals}, we fix $K = 3$ signals of length $L = 21$: see the three red signals in the first column.
\revised{The first signal's actual support has length 11 (rather than 21), which allows us to simulate the situation in which the support of the signal is overestimated.}

Following~\eqref{eq:model}, we generate an observation $y$ of length $12.3 \cdot 10^9$. Each of the three signals appears, respectively (and approximately), $30.0 \cdot 10^6$, $20.0 \cdot 10^6$ and $10.0 \cdot 10^6$ times in $y$ for a total of exactly $60 \cdot 10^6$ occurrences, such that at least $L-1$ zeros separate any two occurrences of any signals according to~\eqref{eq:spacing}. This is done by randomly selecting $60 \cdot 10^6$ placements in $y$, one at a time with an accept/reject rule based on the separation constraint~\eqref{eq:spacing} and locations picked so far. For each placement, one of the three signals is picked at random according to the proportions $\pi = (1/2, 1/3, 1/6)$. Then, i.i.d.\ Gaussian noise with mean zero and standard deviation $\sigma = 3$ is added, to form the observed $y$. 

Visually, the noise dominates the signal to the point that it is challenging to detect occurrences. More precisely, the cross-correlations of $y$ even with the true signals presents peaks at essentially random locations,  uninformative of the actual locations of the signal occurrences. Thus, we contend that it would be difficult for any algorithm to locate the signal occurrences, let alone to cluster them according to which signal appears where.

Our aim is to investigate how accurately we can estimate the signals as a function of the observation length. To this end, we consider a growing part of the observation $y$. For each length, we compute autocorrelations on that part, then we go on to estimate the signals from these ``mixed'' quantities. In practice, the autocorrelations are computed on disjoint segments of $y$ of length $100\cdot10^6$ and added up, without correction for the junction points. Segments are handled sequentially on a GPU, as GPUs are particularly well suited to execute simple instructions across large vectors of data. If multiple GPUs are available, segments can  be handled in parallel.

Having computed the autocorrelations of interest, we estimate signals $x_1, \ldots, x_K$ and coefficients $\gamma_1, \ldots, \gamma_K$ which agree with the data. We choose to do so by running an optimization algorithm on the following nonlinear least-squares problem:
\begin{multline}
	\min_{\substack{\hat x_1, \ldots, \hat x_K \in \R^{W} \\ \hat \gamma_1, \ldots, \hat \gamma_K > 0}} w_1 \left( a_y^1 - \sum_{k=1}^K \hat \gamma_k a_{\hat x_k}^1 \right)^2 + w_2 \sum_{\ell = 1}^{L-1} \left( a_y^2[\ell] - \sum_{k=1}^K \hat \gamma_k a_{\hat x_k}^2[\ell] \right)^2 + \\ w_3 \sum_{\substack{2\leq\ell_1\leq L-1 \\ 1 \leq \ell_2 \leq \ell_1-1}} \left( a_y^3[\ell_1, \ell_2] - \sum_{k=1}^K \hat \gamma_k a_{\hat x_k}^3[\ell_1,\ell_2] \right)^2,
	\label{eq:optim1D}
\end{multline}
where $W \geq L$ is the length of the sought signals and the weights are set to $w_1 = 1/2, w_2 = 1/2n_2, w_3 = 1/2n_3$, where $n_2, n_3$ are the number of coefficients used for each autocorrelation order: $n_2 = L-1$, $n_3 = \frac{(L-1)(L-2)}{2}$ (weights could also be set in accordance with variance estimates as in~\cite{boumal2017heterogeneous}).

Setting $W = L$ (as is a priori desired) is problematic because the above optimization problem appears to have numerous poor local optimizers. Thus, we first run the optimization with $W = 2L-1$. This problem appears to have few poor local optima, perhaps because the additional degrees of freedom allow for more escape directions. Since we hope the signals estimated this way correspond to the true signals zero-padded on either side to length $W$, we extract from each one a subsignal of length $L$ that has largest $\ell_2$-norm. This estimator is then used as initial iterate for~\eqref{eq:optim1D}, this time with $W = L$. We find that this procedure is reliable for a wide range of experimental parameters. To solve~\eqref{eq:optim1D}, we run the trust-region method implemented in Manopt~\cite{manopt} \revised{from 10 different random initial guesses and keep the one with lowest cost function value.} 
\revised{Manopt} allows to treat the positivity constraints on coefficients $\hat \gamma_k$. Notice that the cost function is a polynomial in the variables, so that it is straightforward to compute it and its derivatives.

\revised{To asses reconstruction quality, we report the relative root mean squared error between the estimated signals and the ground truth (up to permutation of the $K$ signals.) For the first signal (with the overestimated support), we also translationally aligned the estimate with the ground truth because its estimation is only possible up to shift.}

In Figure~\ref{fig:1Dheterosignals}, we find that the signals can be recovered with good accuracy despite the noise levels which seemingly hinders location and clustering. We also note that the amount of data required to produce these good estimations is large. Furthermore, as illustrated here and as we have observed in numerous experiments, signals with more variations (such as the third signal in this experiment which was generated once from a Gaussian distribution) are easier to estimate accurately than more regular signals (in this case, despite the fact that the third signal occurs less frequently than the others).
This phenomenon has been also observed in multi-reference alignment~\cite[Section 3.2]{perry2017sample}.

\subsection{Experiment 2}

\begin{figure}[t]
	\centering
	\includegraphics[width=.7\linewidth]{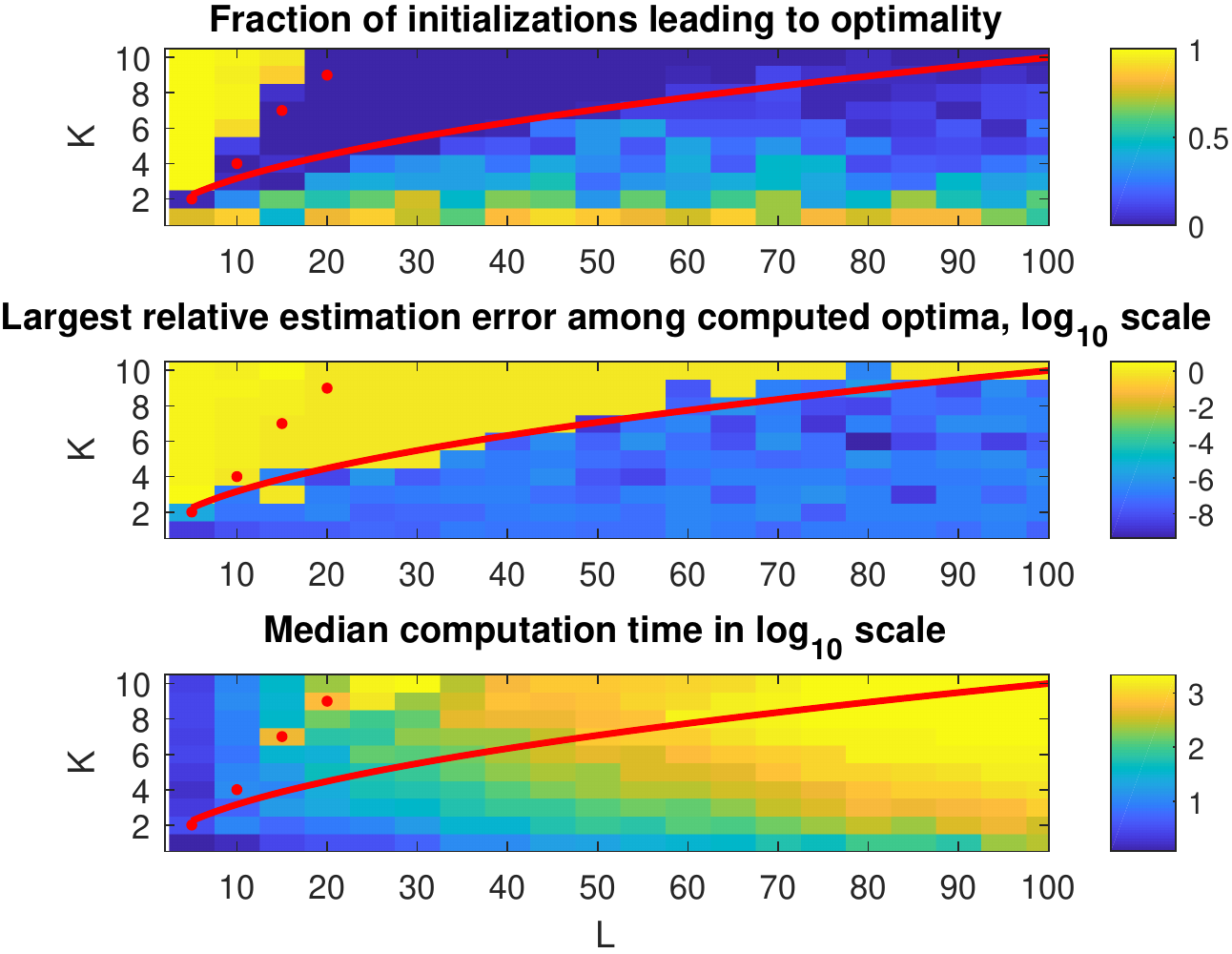}
	\caption{In the $N \to \infty$ regime (access to exact autocorrelations, excluding biased entries) and with known uniform densities, it seems that $K$ up to $\sqrt{L}$ (red curve) i.i.d.\ Gaussian signals of length $L$ can be recovered from the known moments. CPU time is in seconds. Strictly above red dots, recovery is impossible because the number of unknowns exceeds the number of computed autocorrelations; see Section~\ref{sec:heterogeneity}. Similarly to~\cite[Fig.~4.1]{boumal2017heterogeneous}, this experiment suggests a possible statistical-computational gap.}
	\label{fig:KLXP}
\end{figure}

In this second experiment, presented in Figure~\ref{fig:KLXP}, we investigate how many distinct signals ${x_1, \ldots, x_K}$ can be estimated from mixed autocorrelations~\eqref{eq:mixedautocorr}. In order to do so, we consider a setup where the mixed autocorrelations are known perfectly. This corresponds to the limit of an infinitely long observation $y$ with fixed density $\gamma$ and fixed noise level (that may be arbitrarily high). The specific value of $\sigma$ is immaterial since we only consider autocorrelations that are unaffected by noise bias. Furthermore, we assume uniform occurrence distribution $\gamma_k$ (known to the algorithm), and known density $\gamma$ (which is then irrelevant as it only induces a global scaling of the autocorrelations of~$y$).

To produce Figure~\ref{fig:KLXP}, we consider each pair $(K, L)$ in turn, with $K = 1, 2, 3, \ldots, 10$ and $L = 5, 10, 15, \ldots, 100$. For each pair, we generate $K$ random normal signals of length $L$, once. The perfect mixed autocorrelations are computed. They are then provided to the inversion algorithm described in Section~\ref{sec:XP1}, together with the knowledge that signals occur with equal probability, as well as the correct density $\gamma$. The algorithm is initialized 50 times with an independent random initial guess, also following a normal distribution. For each run, we record three metrics:
\begin{enumerate}
	\item Whether the optimization algorithm managed to produce a solution with cost function value below $10^{-16}$: this assesses whether optimization succeeded.
	\item The relative root mean squared error between the estimated signals and the ground truth (up to permutation of the $K$ signals.)
	\item The computation time in seconds (keeping in mind that the 50 runs are done in parallel on the same, shared computer, so that this is more of a qualitative assessment.)
\end{enumerate}
These metrics are summarized and presented in Figure~\ref{fig:KLXP} as three panels.
\begin{enumerate}
	\item Panel 1 shows for each pair $(K, L)$ which fraction of the 50 runs reached optimality (between 0 and 1).
	\item For each pair $(K, L)$, any estimator produced by the optimization algorithm such that the cost function value is close to zero must be considered a valid estimator, since it agrees almost perfectly with the data. For all of those, we compute the error compared to the ground truth. Panel 2 shows the largest such error, on a log scale in base 10. (If optimization never succeeded for that pair, we report a relative error of 1.) A large value means that, among all near global optima of the optimization problem (if any), at least one was a poor estimator. A small value indicates all computed near global optima were good estimators.
	\item Median $\log_{10}$ computation time (where the median is computed after taking the log of the CPU times, in base 10.)
\end{enumerate}

Following~\cite{boumal2017heterogeneous}, overlaid on the panels we trace the red curve $K = \sqrt{L}$ as well as red dots which are computed from the considerations in Section~\ref{sec:heterogeneity} (adapted to the fact that the $\gamma_k$'s are known). We observe that strictly above the red dots the optimization problem appears to be easy to solve (despite non-convexity), yet, as predicted, the corresponding estimators are not informative since there is not enough information in the computed quantities compared to the number of parameters. On the other hand, below the (empirical) red curve, the optimization problem is sometimes solved to optimality (although it may take more than one random initialization to get one successful run), and the corresponding estimators are accurate. In between the red curve and the red dots, the optimization problem appears to be particularly challenging: we essentially never produce a global optimum, hence we also do not have an estimator. This experiment suggests a possible computational-statistical gap in the area between the red curve and the red dots, where it is possible that the signals could be estimated, but perhaps not with a computationally efficient procedure.
Similar results were observed for the multi-reference problem~\cite{boumal2017heterogeneous,weinthesis,bandeira2017estimation}.

\subsection{Experiment 3}

\begin{figure}[t]
	\centering
	\begin{subfigure}[h]{0.33\linewidth}
		\centering
		\includegraphics[width=.8\linewidth]{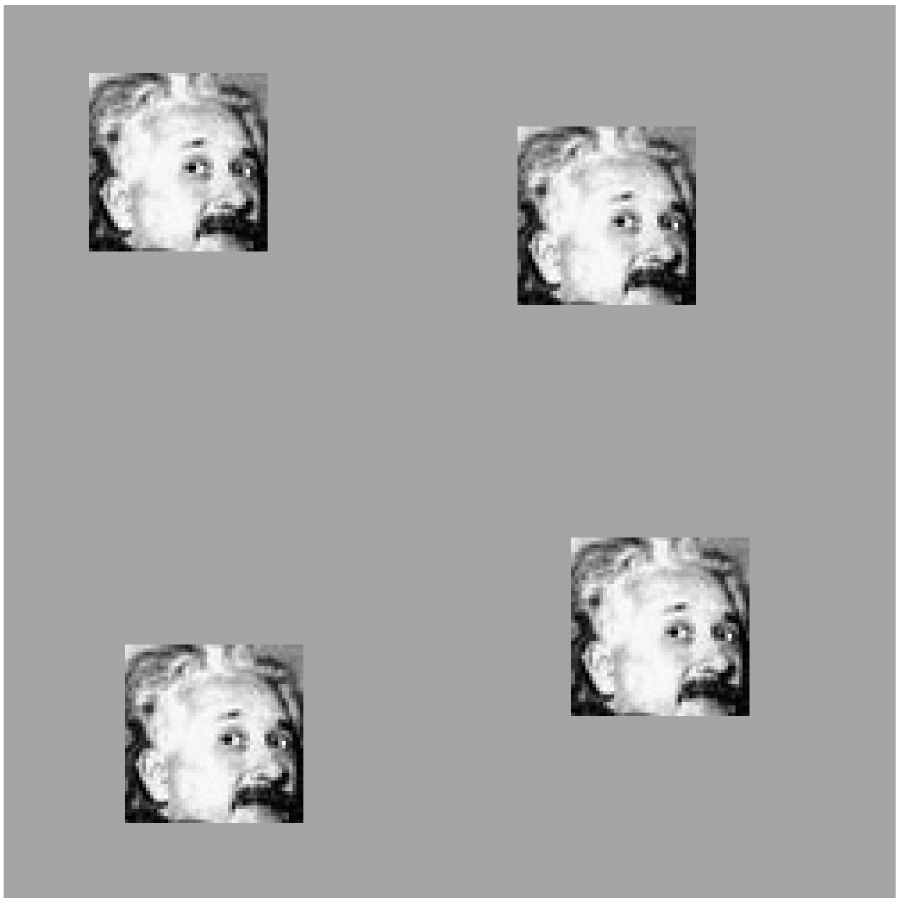}
		\caption{$\sigma = 0$}
	\end{subfigure}%
	\begin{subfigure}[h]{0.33\linewidth}
		\centering
		\includegraphics[width=.8\linewidth]{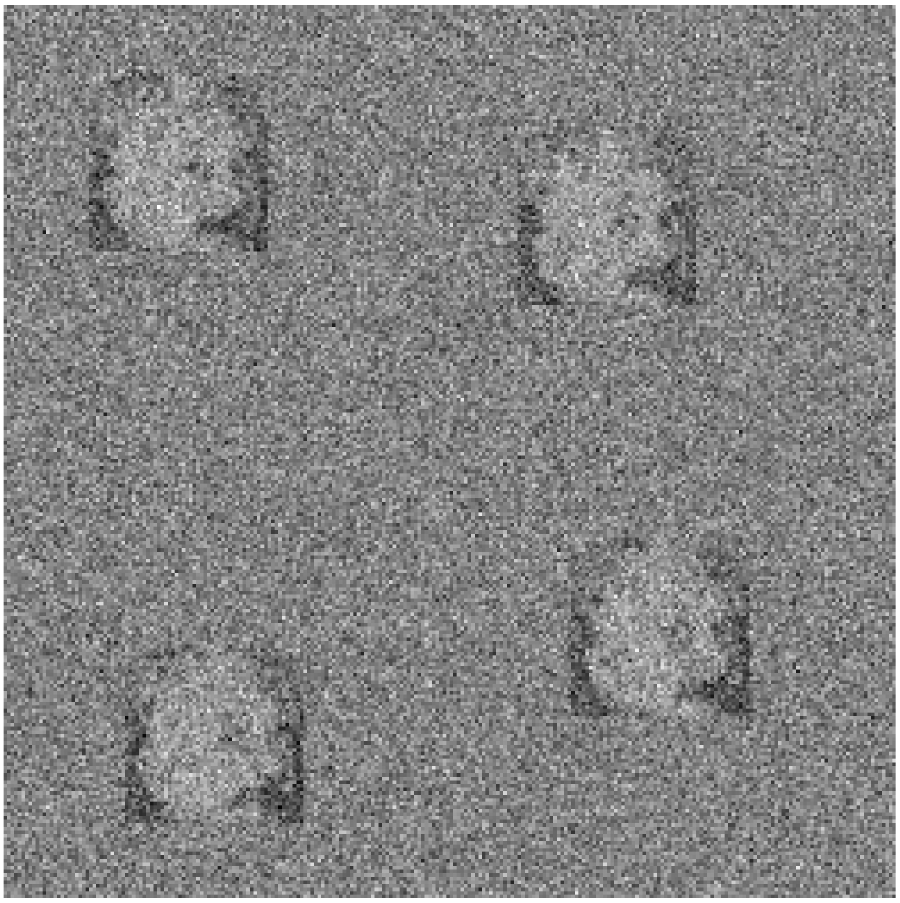}
		\caption{$\sigma = 0.5$}
	\end{subfigure}
	\begin{subfigure}[h]{0.33\linewidth}
		\centering
		\includegraphics[width=.8\linewidth]{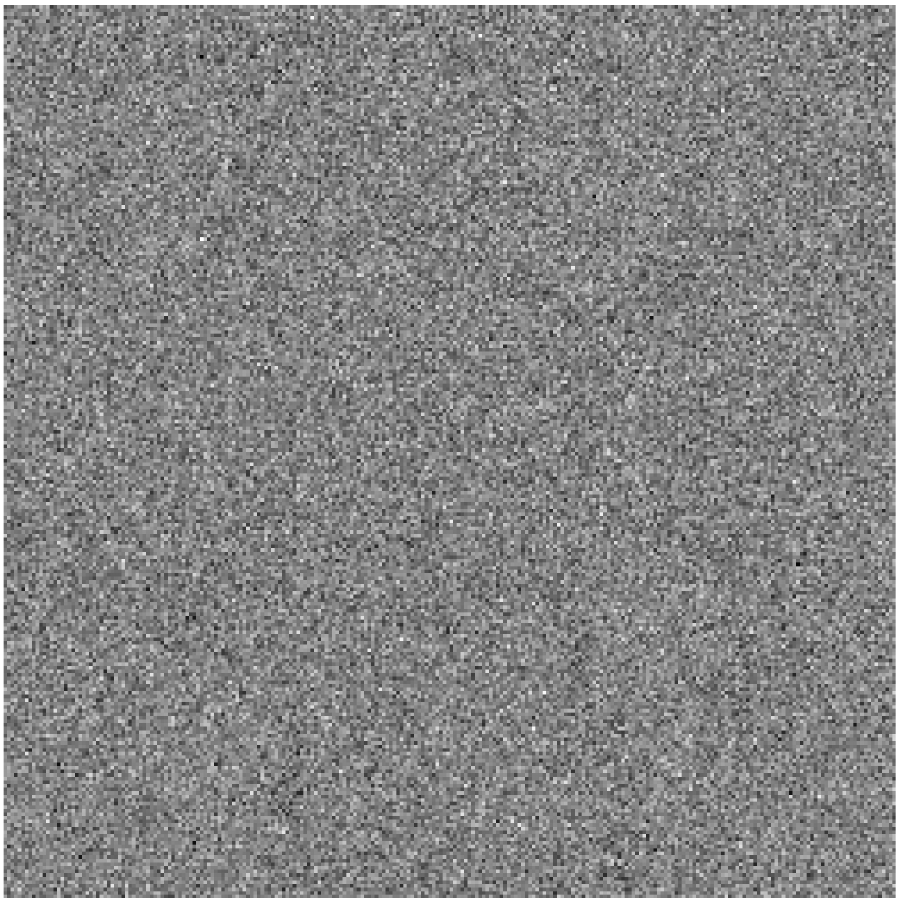}
		\caption{$\sigma = 3$}
	\end{subfigure}
	\caption{\label{fig:micro_example} Example of observations for the 2-D experiment (of size $250\times 250$) with additive white Gaussian noise of variance $\sigma^2$ for increasing values of $\sigma$. Each observation contains the same four occurrences of a $50 \times 50$ image of Einstein. In panel (c), the noise level is such that it is challenging to detect  the planted images.} 
\end{figure}

Autocorrelation analysis can be carried out in dimensions greater than one. In the following experiment, we estimate 
a 50-by-50 pixel grayscale picture of Einstein with mean zero from a growing number of observations $y$. Each observation is of size $4096\times 4096$ pixels and contains 700 occurrences on average at random locations, while maintaining the separation condition~\eqref{eq:spacing} in each axis separately.
The observations are contaminated with additive white Gaussian noise with standard deviation $\sigma = 3$, illustrated in Figure~\ref{fig:micro_example}. 

We compute the average second-order autocorrelation of the observations. This is a particularly simple computation which can be efficiently executed with a fast Fourier transform (FFT), in parallel over the numerous observations. We assume the number of signal occurrences (akin to the density $\gamma$) and the standard deviation of the noise, $\sigma$, are known. Given those quantities, the second-order autocorrelation of the image can be easily deduced from~\eqref{eq:ac2_micrograph}. As explained in Section~\ref{sec:high_dimensions}, an image is determined uniquely form its second-order autocorrelations. Then, to estimate the target image, we use a standard phase retrieval algorithm called relaxed-reflect-reflect (RRR)~\cite{elser2017rrr}, initialized randomly.

\begin{figure}[t]
	\centering
	\includegraphics[width=1\linewidth]{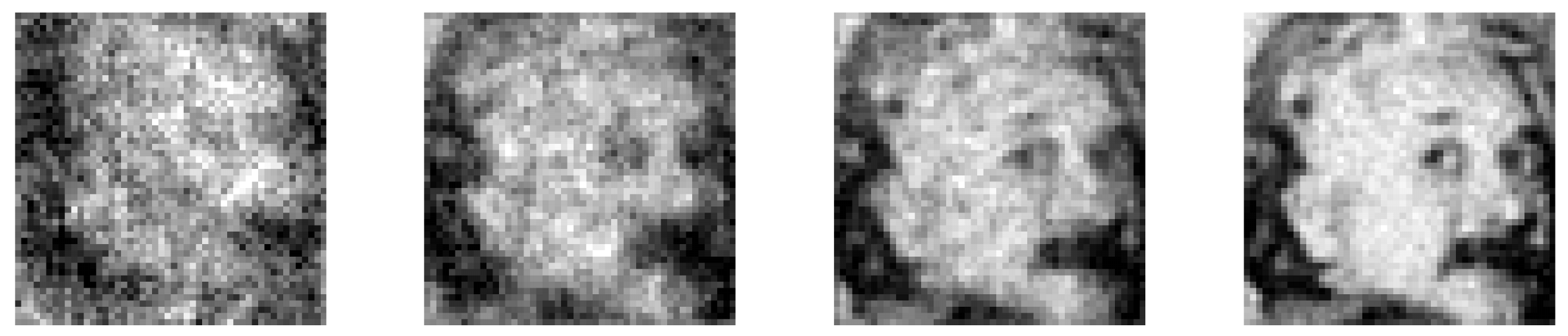}
	\caption{\label{fig:Einst_example} Recovery of Einstein from observations at noise level $\sigma = 3$ (see Figure~\ref{fig:micro_example}(c)). Averaged autocorrelations of the data allow to estimate the power spectrum of the target image. This does not require locating the signal occurrences. The RRR algorithm produces the estimates,  obtained from $2\times 10^2,2\times 10^3,2\times 10^4$ and $2\times 10^5$ observations (growing across panels from left to right).}	
\end{figure}

\begin{figure}[h]
	\centering
	\includegraphics[width=.8\linewidth]{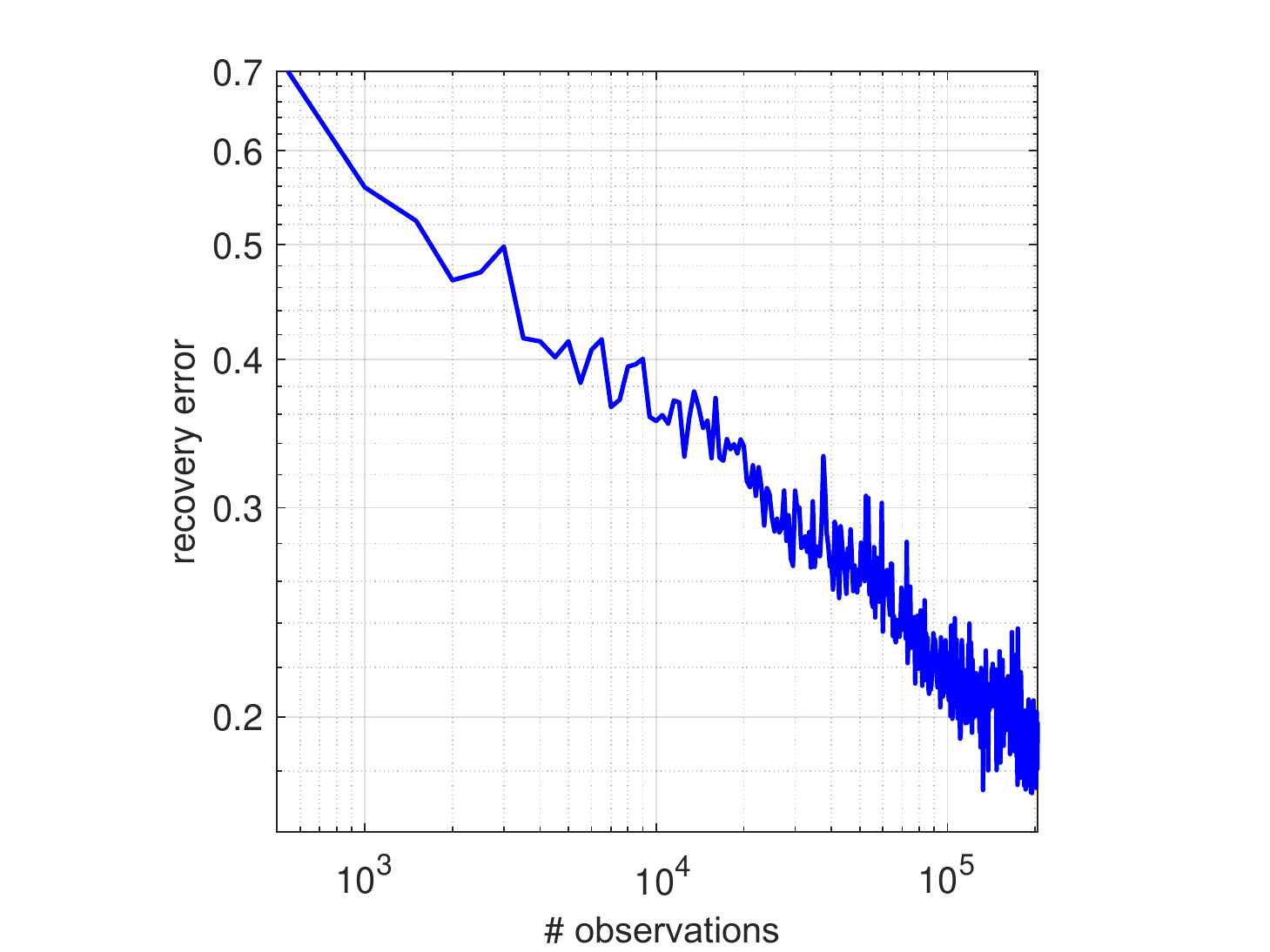}
	\caption{\label{fig:error_per_micro}Relative error curve for Experiment 3 in Figure~\ref{fig:Einst_example}. Each observation contains about 700 image occurrences at unknown locations.}
\end{figure}

Relative error is measured as the ratio of the root mean square error to the norm of the ground truth (square root of the sum of squared pixel intensities). Figure~\ref{fig:Einst_example} shows several estimated images for a growing number of observations. Figure~\ref{fig:error_per_micro} presents the normalized recovery error as a function of the amount of data available.  This is computed after fixing the reflection symmetries (see Section~\ref{sec:high_dimensions}).

As evidenced by these figures, the ground truth image can be estimated increasingly well from increasingly many observations, without the need to locate the signal occurrences.

\section{Summary}

This paper suggests a computational  framework for estimation under extreme noise levels. 
The crux of the method lies in the distinction between parameters of interest (the signals) and nuisance variables (parameters associated with individual signal occurrences, such as location and class). In part through theory and in part through numerical experiments, we show that estimating the signals is possible even when they cannot be detected in the data. The method consists of two steps. First, we estimate the autocorrelations of the observation. A key feature is that, for any noise level, these autocorrelations can be estimated to any desired accuracy given sufficiently rich observations. Second, we recover the signals from the autocorrelations. This recovery entails solving a system of low-order polynomial equations. While solving such systems is hard in general, we found that in the homogeneous case we can solve them explicitly, and in both the homogeneous and heterogeneous cases we can solve them with reasonable robustness through non-convex optimization, in a wide regime of parameters.

\revised{In addition, autocorrelation analysis provides a flexible framework to extend the multi-target detection model by  relating the expected autocorrelations
of the data with the signals, and all parameters necessary to describe the generative model. For instance,  a follow-up paper relaxes the separation condition~\eqref{eq:spacing} and allows  an arbitrary spacing of targets, as long as the signal occurrences do not overlap~\cite{lan2019multi}. In that case, the autocorrelations of the data are functions of the signal and the unknown target distribution. In a similar fashion, one may include more realistic noise models, the effect of a blurring kernel (e.g., the point spread function of the microscope) and so on.}

The prime motivation of this paper emanates from challenges in small particle reconstruction using cryo-EM.
Small particles induce such low signal-to-noise ratio in the micrograph that particle picking---the first step in any current cryo-EM reconstruction algorithm---is impossible. 
The main message of our recent report~\cite{bendory2018toward} is that particle picking is merely a means to an end (although admittedly of key usefulness when it can be done): the locations and classes of individual particle projections are nuisance variables. The ultimate goal is only to estimate the 3-D structures. 
To this end,  we used autocorrelation analysis to estimate the structure directly from the micrograph, without particle picking. 
In order to gain better understanding of the method, this paper focuses on an abstraction of cryo-EM---the multi-target detection model. 
Our next goal is to reconsider the full cryo-EM model, both from theoretical and algorithmic perspectives. 
In particular, the numerical results in~\cite{bendory2018toward} suggest that the achieved resolution using autocorrelations up to third order is limited by ill-conditioning of the system of polynomial equations. Higher resolution may require computing higher-order autocorrelations, which would increase the sample complexity and computational complexity of the algorithm. Despite the challenges, we believe that this approach may ultimately offer a way to reconstruct 3-D structures that are too small for current algorithmic pipelines.

%

\section*{Acknowledgment} 
The authors  thank Ayelet Heimowitz, Joe Kileel, Ti-Yen Lan and Amit Moscovich  for helpful discussions.
The research is supported in parts by Award Number R01GM090200 from the NIGMS, FA9550-17-1-0291 from AFOSR, Simons Foundation Math+X Investigator Award, the Moore Foundation Data-Driven Discovery Investigator Award, and NSF BIGDATA Award IIS-1837992.
NB is partially supported by NSF award DMS-1719558.

\bibliographystyle{plain}

\appendix

\section{Autocorrelations in the well-separated model} \label{sec:autocorrelation_computation}

Let $x_{(1)}, \ldots, x_{(|s|)}$ denote the (independent) realizations of the random signal $x$ in the observation $y$, starting at (deterministic) positions $s_{(1)}, \ldots, s_{(|s|)}$. Let $I_{ij}$ be the indicator variable for whether position $i$ is in the support of occurrence $j$, that is, it is one if $i$ is in $\{s_{(j)}, \ldots, s_{(j)}+L-1\}$, and zero otherwise. Then,
\begin{align}
y[i] & = \sum_{j = 1}^{|s|} I_{ij} x_{(j)}[i-s_{(j)}] + \varepsilon[i].
\label{eq:explicityiindicators}
\end{align}
This gives a simple expression for the first autocorrelation of $y$. Indeed,
\begin{align}
a_y^1 & = \E_y\left\{ \frac{1}{N} \sum_{i = 0}^{N-1} y[i] \right\} \\
& = \frac{1}{N} \E_{x_{(1)}, \ldots, x_{(|s|)}, \varepsilon}\left\{ \sum_{i = 0}^{N-1} \sum_{j = 1}^{|s|} I_{ij} x_{(j)}[i-s_{(j)}] + \varepsilon[i] \right\}.
\end{align}
Now switch the sums over $i$ and $j$, and observe that $I_{ij}$ is zero unless $i = s_{(j)} + t$ for $t$ in the range $0, \ldots, L-1$. Hence,
\begin{align}
a_y^1 & = \frac{1}{N} \sum_{j = 1}^{|s|} \E_{x_{(j)}}\left\{ \sum_{t = 0}^{L-1} x_{(j)}[t]\right\} + \frac{1}{N} \E_\varepsilon\left\{ \sum_{i=0}^{N-1} \varepsilon[i]\right\}.
\end{align}
Since the noise has zero mean and $x_{(1)}, \ldots, x_{(|s|)}$ are identically distributed, we further find:
\begin{align}
a_y^1 & = \frac{|s|L}{N} a_x^1 = \gamma a_x^1.
\end{align}

To address the second-order moments, we resort to the separation condition~\eqref{eq:spacing}. First, consider this expression:
\begin{align*}
N \cdot a_y^2[\ell] & = \E_y\left\{ \sum_{i = 0}^{N-\ell-1} y[i] y[i+\ell] \right\} \\
& = \sum_{i = 0}^{N-\ell-1} \E_{x_{(1)}, \ldots, x_{(|s|)}, \varepsilon}\Bigg\{ \left( \sum_{j = 1}^{|s|} I_{ij} x_{(j)}[i-s_{(j)}] + \varepsilon[i] \right) \cdot \\
& \qquad \qquad \qquad \qquad \qquad  \left( \sum_{j' = 1}^{|s|} I_{i+\ell,j'} x_{(j')}[i+\ell-s_{(j')}] + \varepsilon[i+\ell] \right)  \Bigg\} \\
& = \sum_{i = 0}^{N-\ell-1} \E_{x_{(1)}, \ldots, x_{(|s|), \varepsilon}}\Bigg\{ \sum_{j = 1}^{|s|} \sum_{j' = 1}^{|s|} I_{ij}  I_{i+\ell,j'} x_{(j)}[i-s_{(j)}]  x_{(j')}[i+\ell-s_{(j')}] \\
& \qquad \qquad \qquad \qquad \qquad \qquad + \sum_{j = 1}^{|s|} I_{ij} x_{(j)}[i-s_{(j)}] \varepsilon[i+\ell] \\
& \qquad \qquad \qquad \qquad \qquad \qquad + \sum_{j' = 1}^{|s|} I_{i+\ell,j'} x_{(j')}[i+\ell-s_{(j')}] \varepsilon[i] \\
& \qquad \qquad \qquad \qquad \qquad \qquad + \varepsilon[i] \varepsilon[i + \ell] \Bigg\}.
\end{align*}
The cross-terms vanish in expectation since $\varepsilon$ is zero mean and independent from the signal occurrences. The last term vanishes in expectation unless $\ell = 0$ since distinct entries of $\varepsilon$ are independent. For $\ell = 0$, $\E\{\varepsilon[i]^2\} = \sigma^2$. Finally, using the separation property, observe that if $I_{ij}  I_{i+\ell,j'}$ is nonzero, then it is equal to one, $j = j'$ and $i = s_{(j)} + t$ for some $t$ in $0, \ldots, L-\ell-1$. Then, switch the order of summations to get
\begin{align}
N \cdot a_y^2[\ell] & = \sum_{j=1}^{|s|} \E_{x_{(j)}}\left\{ \sum_{t = 0}^{L-\ell-1} x_{(j)}[t] x_{(j)}[t+\ell] \right\} + (N-\ell)\sigma^2 \delta[\ell],
\end{align}
where $\delta[0] = 1$ and $\delta[\ell \neq 0] = 0$. Since each $x_{(j)}$ is distributed as $x$, they all have the same autocorrelations as $x$ and we finally get
\begin{align}
a_y^2[\ell] & = \gamma a_x^2[\ell] + \frac{N-\ell}{N}\sigma^2 \delta[\ell] = \gamma a_x^2[\ell] + \sigma^2 \delta[\ell].
\end{align}

We now turn to the third-order autocorrelations. These involve the sum
\begin{align}
\sum_{i=0}^{N-\max(\ell_1, \ell_2)-1} y[i] y[i+\ell_1] y[i+\ell_2].
\end{align}
Using~\eqref{eq:explicityiindicators}, we find that this quantity can be expressed as a sum of eight terms:
\begin{enumerate}
	\item $\sum_i \sum_{j,j',j'' = 1}^{|s|} I_{ij} I_{i+\ell_1, j'} I_{i+\ell_2,j''} x_{(j)}[i-s_{(j)}] x_{(j')}[i+\ell_1-s_{(j')}] x_{(j'')}[i+\ell_2-s_{(j'')}]$
	\item $\sum_i \sum_{j,j' = 1}^{|s|} I_{ij} I_{i+\ell_1, j'} x_{(j)}[i-s_{(j)}] x_{(j')}[i+\ell_1-s_{(j')}] \varepsilon[i+\ell_2]$
	\item $\sum_i \sum_{j,j'' = 1}^{|s|} I_{ij} I_{i+\ell_2,j''} x_{(j)}[i-s_{(j)}] \varepsilon[i+\ell_1] x_{(j'')}[i+\ell_2-s_{(j'')}]$
	\item $\sum_i \sum_{j',j'' = 1}^{|s|} I_{i+\ell_1, j'} I_{i+\ell_2,j''} \varepsilon[i] x_{(j')}[i+\ell_1-s_{(j')}] x_{(j'')}[i+\ell_2-s_{(j'')}]$
	\item $\sum_i \sum_{j = 1}^{|s|} I_{ij} x_{(j)}[i-s_{(j)}] \varepsilon[i+\ell_1] \varepsilon[i+\ell_2]$
	\item $\sum_i \sum_{j' = 1}^{|s|} I_{i+\ell_1, j'} \varepsilon[i] x_{(j')}[i+\ell_1-s_{(j')}] \varepsilon[i+\ell_2]$
	\item $\sum_i \sum_{j'' = 1}^{|s|} I_{i+\ell_2,j''} \varepsilon[i] \varepsilon[i+\ell_1] x_{(j'')}[i+\ell_2-s_{(j'')}]$
	\item $\sum_i \varepsilon[i] \varepsilon[i+\ell_1] \varepsilon[i+\ell_2]$
\end{enumerate}
Terms 2--4 and 8 vanish in expectation since odd moments of centered Gaussian variables are zero. For the first term, we use the fact that the separation condition implies
\begin{multline}
I_{ij} I_{i+\ell_1, j'} I_{i+\ell_2,j''} = 1 \iff \\ j=j'=j'' \textrm{ and } i = s_{(j)} + t \textrm{ with } t \in \{ 0, \ldots L-\max(\ell_1, \ell_2)-1 \}.
\end{multline}
(Otherwise, the product of indicators is zero.) This allows to reduce the summations over $j,j',j''$ to a single sum over $j$. Then, switching the order of summation with $i$, we get that the first term is equal to
\begin{align}
\sum_{j=1}^{|s|} \sum_{t=0}^{L-\max(\ell_1, \ell_2)-1} x_{(j)}[t] x_{(j)}[t+\ell_1] x_{(j)}[t+\ell_2].
\end{align}
In expectation over the realizations $x_{(j)}$, using again that they are i.i.d.\ with the same distribution as $x$, this first term yields $|s|L a_x^3[\ell_1, \ell_2]$. Now consider the fifth term. Taking expectation against $\varepsilon$ yields
\begin{align}
\sum_{i=0}^{N-\max(\ell_1, \ell_2)-1} \sum_{j = 1}^{|s|} I_{ij} x_{(j)}[i-s_{(j)}] \sigma^2 \delta[\ell_1 - \ell_2].
\end{align}
Switch the order of summation over $i$ and $j$ again to get
\begin{align}
\sigma^2 \delta[\ell_1 - \ell_2] \sum_{j = 1}^{|s|} \sum_{t=0}^{L-1} x_{(j)}[t].
\end{align}
Now taking expectation against the signal occurrences yields $|s|L \sigma^2 a_x^1 \delta[\ell_1 - \ell_2]$. A similar reasoning for terms 6 and 7 yields this final formula for the third-order autocorrelations of $y$:
\begin{align}
a_y^3[\ell_1, \ell_2] & = \gamma a_x^3[\ell_1, \ell_2] + \gamma \sigma^2 a_x^1 \left( \delta[\ell_1] + \delta[\ell_2] + \delta[\ell_1 - \ell_2] \right).
\end{align}


\section{Autocorrelations in the Poisson model} \label{sec:proof_prop_poisson}

We will denote by $m_l$ the moment tensors of $x$:
\begin{align}
m_1[i] = \E x[i], \quad 0 \le i \le L-1,
\end{align}
\begin{align}
m_2[i,j] = \E x[i] x[j], \quad 0 \le i,j \le L-1,
\end{align}
\begin{align}
m_3[i,j,k] = \E x[i] x[j] x[k], \quad 0 \le i,j,k \le L-1.
\end{align}

We obtain the autocorrelations $a_x^l$ of $x$ by averaging over a slice of the moment tensors:
\begin{align}
a_x^1 = \frac{1}{L}\sum_{i=0}^{L-1} m_1[i],
\end{align}
\begin{align}
a_x^2[\ell] = \frac{1}{L} \sum_{i=0}^{L-1} m_2[i,i+\ell],
\end{align}
and
\begin{align}
a_x^3[\ell_1,\ell_2] = \frac{1}{L}\sum_{i=0}^{L-1} m_3[i,i+\ell_1,i+\ell_2].
\end{align}

We will make repeated use of the following elementary lemma:
\begin{lem} \label{lem-choose}
	If $z \sim \Poisson(\lambda)$, then 
	\begin{align}
	\E {z\choose k} = \frac{\lambda^k}{k!}.
	\end{align}
\end{lem}

\subsection{Computing $a_y^1$}

We will first condition on the vector $s$ of locations of the subsignals in $y$, and then average over $s$. We will denote by $x_{(1)}^{i},\dots,x_{(s[i])}^i$ the random vectors starting  in $y[i]$. We have:
%
\begin{align}
\E[ y[i] | s ] = \sum_{j=0}^{L-1} \sum_{k=1}^{s[i-j]} \E x_{(k)}^{i-j}[j]
= \sum_{k=1}^{s[i-j]} L a_x^1
= s[i-j] L a_x^1.
\end{align}
Now taking expectations over $s$ and using $\E s[i-j] = \gamma/L$ we get:
\begin{align}
\E y[i] = \gamma a_x^1.
\end{align}
Consequently,
\begin{align}
a_y^1 = \frac{1}{N} \sum_{i=1}^n \E y[i] = \gamma a_x^1.
\end{align}

\subsection{Computing $a_y^2$}

First we will consider the noise-free case, where $\sigma = 0$. We will condition on $s$ first, and then take the expectation over $s$. Fix $i_1 \ne i_2$, and let $\ell = i_2 - i_1$. Then:
\begin{align}
y[i_1] y[i_2]
&= \sum_{j_1=0}^{L-1} \sum_{j_2=0}^{L-1} 
\sum_{k_1=1}^{s[i_1-j_1]}\sum_{k_2=1}^{s[i_2-j_2]}
x_{(k_1)}^{i_1-j_1}[j_1] x_{(k_2)}^{i_2 - j_2}[j_2].
\end{align}

We break up the double sum over $j_1$ and $j_2$ into two terms: one where $j_2 \ne j_1 + \ell$, and one where $j_2 = j_1 + \ell$ or equivalently $i_1-j_1 = i_2-j_2$. In the first case, all the terms are independent, and so the expectation factors. In the second case, when $k_1 \ne k_2$ we have independence, but otherwise not. This gives (all expectations are conditional on $s$):
\begin{align} \label{moment2-condm}
\E y[i_1] y[i_2]
=& \sum_{j_1=0}^{L-1} \sum_{j_2=0}^{L-1} 
\sum_{k_1=1}^{s[i_1-j_1]}\sum_{k_2=1}^{s[i_2-j_2]}
\E x_{(k_1)}^{i_1-j_1}[j_1] x_{(k_2)}^{i_2 - j_2}[j_2]
\nonumber \\
=& \sum_{j_1 - j_2 \ne \ell} \sum_{k_1} 
\sum_{k_2} \E x_{(k_1)}^{i_1-j_1}[j_1] x_{(k_2)}^{i_2 - j_2}[j_2]
\nonumber \\
& + \sum_{j_1 = 0}^{L-1} \sum_{k_1 \ne k_2} 
\E x_{(k_1)}^{i_1-j_1}[j_1] x_{(k_2)}^{i_1 - j_1}[j_1+\ell]
\nonumber \\
& + \sum_{j_1 = 0}^{L-1} \sum_{k_1=1}^{s[i_1-j_1]} 
\E x_{(k_1)}^{i_1-j_1}[j_1] x_{(k_1)}^{i_1-j_1}[j_1 + \ell] 
\nonumber \\
=& \sum_{j_1 - j_2 \ne \ell} s[i_1-j_1] s[i_2 - j_2] m_1[j_1] m_1[j_2]
\nonumber \\
& + \sum_{j_1 = 0}^{L-1} s[i_1-j_1](s[i_1-j_1] - 1) m_1[j_1] m_1[j_1 + \ell]
\nonumber \\
& + \sum_{j_1 = 0}^{L-1} s[i_1-j_1] m_2[j_1,j_1 + \ell] .
\end{align}
Now take expectations over the Poisson random variables, using Lemma \ref{lem-choose}:
\begin{align}
\E y[i_1] y[i_2]
=& \sum_{j_1 - j_2 \ne \ell} \E s[i_1-j_1] s[i_2 - j_2] m_1[j_1] m_1[j_2]
\nonumber \\
&       + \sum_{j_1 = 0}^{L-1} \E s[i_1-j_1](s[i_1-j_1] - 1) m_1[j_1] m_1[j_1 + \ell]
\nonumber \\
&       + \sum_{j_1 = 0}^{L-1} \E s[i_1-j_1] m_2[j_1,j_1 + \ell]
\nonumber \\
=& \frac{1}{L^2}\sum_{j_1 - j_2 \ne \ell} \gamma^2 m_1[j_1] m_1[j_2]
+ \frac{1}{L^2} \sum_{j_1 = 0}^{L-1} \gamma^2 m_1[j_1] m_1[j_1 + \ell]
\nonumber \\
&       + \frac{1}{L}\sum_{j_1 = 0}^{L-1} \gamma m_2[j_1,j_1 + \ell]
\nonumber \\
=&  \bigg(\frac{\gamma}{L}  \sum_{j = 0}^{L-1} m_1[j] \bigg)^2
+ \frac{\gamma}{L} \sum_{j = 0}^{L-1} m_2[j,j + \ell]
\nonumber \\
=&  (\gamma a_x^1)^2 + \gamma a_x^2[\ell].
\end{align}
For positive $\sigma$, we observe that any terms linear in the noise vanish in expectation. Denoting by $x^*$ the clean signal component of length $N$, so $y = x^* + \ep$, we have:
\begin{align}
\E y[i_1] y[i_2] = \E x^*[i_1] x^*[i_2] + \E \ep[i_1] \ep[i_2]
= (\gamma a_x^1)^2 + \gamma a_x^2[\ell] + \sigma^2 \delta[i_1 - i_2].
\end{align}
We conclude by averaging over $i_1$ and $i_2$ with a fixed value of $\ell = i_2 - i_1$.

\subsection{Computing $a_x^3$}

We will first assume $\sigma = 0$. For three distinct $i_1$, $i_2$ and $i_3$, we let $\ell_1 = i_2 - i_1$ and $\ell_2 = i_3 - i_1$. We have:
\begin{align}
& y[i_1] y[i_2] y[i_3]
= \sum_{j_1=0}^{L-1} \sum_{j_2=0}^{L-1} \sum_{j_3=0}^{L-1} 
\sum_{k_1=1}^{s[i_1-j_1]}\sum_{k_2=1}^{s[i_2-j_2]} \sum_{k_3=1}^{s[i_3-j_3]}
x_{(k_1)}^{i_1-j_1}[j_1] x_{(k_2)}^{i_2 - j_2}[j_2] x_{(k_3)}^{i_3 - j_3}[j_3].
\end{align}
We will break up the outer three sums into disjoint sums with the following ranges of indices:
\begin{enumerate}
	
	\item \label{case1}
	$j_2 = j_1 + \ell_1$ and $j_3 = j_2 + \ell_2 - \ell_1$.
	
	\item \label{case2}
	$j_2 = j_1 + \ell_1$ and $j_3 \ne j_2 + \ell_2 - \ell_1$.
	
	\item \label{case3}
	$j_2 \ne j_1 + \ell_1$ and $j_3 = j_1 + \ell_2$.
	
	\item \label{case4}
	$j_2 \ne j_1 + \ell_1$ and $j_3 \ne j_1 + \ell_2$ and $j_3 = j_2 + \ell_2 - \ell_1$.
	
	\item \label{case5}
	$j_2 \ne j_1 + \ell_1$ and $j_3 \ne j_1 + \ell_2$ and $j_3 \ne j_2 + \ell_2 - \ell_1$.
	
\end{enumerate}

For Case \ref{case1}, we have $\ell \equiv i_1 - j_1 = i_2 - j_2 = i_3 - j_3$. We further break up the sum:
\begin{align}
&\sum_{j=0}^{L-1} \sum_{k_1=1}^{s[\ell]} \sum_{k_2=1}^{s[\ell]} \sum_{k_3=1}^{s[\ell]} 
x_{(k_1)}^{\ell}[j] x_{(k_2)}^{\ell}[j + \ell_1] x_{(k_3)}^{\ell}[j + \ell_2]
\nonumber \\
=& \underbrace{ \sum_{j=0}^{L-1} \sum_{k_i \text{distinct}} 
	x_{(k_1)}^{\ell}[j] x_{(k_2)}^{\ell}[j + \ell_1] x_{(k_3)}^{\ell}[j + \ell_2]
}_{\text{(a)}}
\nonumber \\
&+\underbrace{ \sum_{j=0}^{L-1} \sum_{k_1=k_2\ne k_3} 
	x_{(k_1)}^{\ell}[j] x_{(k_2)}^{\ell}[j + \ell_1] x_{(k_3)}^{\ell}[j + \ell_2]
}_{\text{(b)}}
\nonumber \\
&+\underbrace{ \sum_{j=0}^{L-1} \sum_{k_1=k_3\ne k_2} 
	x_{(k_1)}^{\ell}[j] x_{(k_2)}^{\ell}[j + \ell_1] x_{(k_3)}^{\ell}[j + \ell_2]
}_{\text{(c)}}
\nonumber \\
&+\underbrace{ \sum_{j=0}^{L-1} \sum_{k_2=k_3\ne k_1} 
	x_{(k_1)}^{\ell}[j] x_{(k_2)}^{\ell}[j + \ell_1] x_{(k_3)}^{\ell}[j + \ell_2]
}_{\text{(d)}}
\nonumber \\
&+\underbrace{ \sum_{j=0}^{L-1} \sum_{k_1=k_2=k_3} 
	x_{(k_1)}^{\ell}[j] x_{(k_2)}^{\ell}[j + \ell_1] x_{(k_3)}^{\ell}[j + \ell_2]
}_{\text{(e)}}.
\end{align}

For term (a), the expectation conditional on $s$ is:
\begin{align}
\sum_{j=0}^{L-1} s[\ell](s[\ell]-1)(s[\ell]-2)m_1[j] m_1[j+\ell_1] m_1[j+\ell_2].
\end{align}
Using Lemma \ref{lem-choose}, the unconditional expectation of (a) is then:
\begin{align} \label{aaaa}
\frac{\gamma^3}{L^3} \sum_{j=0}^{L-1} m_1[j] m_1[j+\ell_1] m_1[j+\ell_2].
\end{align}

For term (b), the expectation conditional on $s$ is:
\begin{align}
\sum_{j=0}^{L-1} s[\ell] (s[\ell] - 1) m_2[j,j+\ell_1] m_1[j + \ell_2]
\end{align}
and then again using Lemma \ref{lem-choose} we get the expected value:
\begin{align} \label{bbbb}
\frac{\gamma^2}{L^2} \sum_{j=0}^{L-1} m_2[j,j+\ell_1] m_1[j+\ell_2].
\end{align}

Similarly, the expected values of terms (c) and (d) are:
\begin{align} \label{cccc}
\frac{\gamma^2}{L^2} \sum_{j=0}^{L-1} m_2[j,j+\ell_2] m_1[j+\ell_1].
\end{align}
and
\begin{align} \label{dddd}
\frac{\gamma^2}{L^2} \sum_{j=0}^{L-1} m_2[j+\ell_1,j+\ell_2] m_1[j].
\end{align}

Finally, the expected value of term (e) is easily shown to be:
\begin{align} \label{eeee}
\frac{\gamma}{L} \sum_{j=0}^{L-1} m_3[j,j+\ell_1,j+\ell_2].
\end{align}
This concludes the computation for Case \ref{case1}.

Moving onto Case \ref{case2}, we have $\Delta_1 \equiv i_1 - j_1 = i_2 - j_2$, and also define $\Delta_2 \equiv i_3 - j_3$. By definition, $\Delta_1 \ne \Delta_2$. The sum is:
\begin{align}
& \sum_{j_1=0}^{L-1} \sum_{j_3 \ne j_1 + \ell_2}
\sum_{1 \le k_1,k_2 \le s[\Delta_1]} \sum_{k_3=1}^{s[\Delta_2]}
x_{(k_1)}^{\Delta_1}[j_1] x_{(k_2)}^{\Delta_1}[j_1 + \ell_1] x_{(k_3)}^{\Delta_2}[j_3]   
\nonumber \\
=& \sum_{j_1=0}^{L-1} \sum_{j_3 \ne j_1 + \ell_2} \sum_{k_3=1}^{s[\Delta_2]}
\Bigg\{  \sum_{1 \le k_1 \ne k_2 \le s[\Delta_1]} 
x_{(k_1)}^{\Delta_1}[j_1] x_{(k_2)}^{\Delta_1}[j_1 + \ell_1] x_{(k_3)}^{\Delta_2}[j_3]
\nonumber \\
& + \sum_{k_1=1}^{s[\Delta_1]} x_{(k_1)}^{\Delta_1}[j_1] x_{(k_1)}^{\Delta_1}[j_1 + \ell_1] 
x_{(k_3)}^{\Delta_2}[j_3]  \Bigg\}.
\end{align}
Taking expectations conditional on $s$, we then get:
\begin{align}
\sum_{j_1=0}^{L-1} \sum_{j_3 \ne j_1 + \ell_2} 
\Bigg( & s[\Delta_1] (s[\Delta_1]-1) s[\Delta_2] m_1[j_1] m_1[j_1 + \ell_1] m_1[j_3]
\nonumber \\
&  + s[\Delta_1] s[\Delta_2] m_2[j_1,j_1+\ell_1] m_1[j_3] \Bigg).
\end{align}

Taking expectations over $s$ and using Lemma \ref{lem-choose} then gives:
\begin{align}
& \frac{\gamma^3}{L^3} \sum_{j_1=0}^{L-1} \sum_{j_3 \ne j_1 + \ell_2} 
m_1[j_1] m_1[j_1 + \ell_1] m_1[j_3]
\label{ffff} \\
& + \frac{\gamma^2}{L^2} \sum_{j_1=0}^{L-1} \sum_{j_3 \ne j_1 + \ell_2}  
m_2[j_1,j_1+\ell_1] m_1[j_3].
\label{gggg}
\end{align}

Similarly, Cases \ref{case3} and \ref{case4} give the expressions:
\begin{align}
& \frac{\gamma^3}{L^3} \sum_{j_1=0}^{L-1} \sum_{j_2 \ne j_1 + \ell_1} 
m_1[j_1] m_1[j_1 + \ell_2] m_1[j_2]
\label{hhhh} \\
& + \frac{\gamma^2}{L^2} \sum_{j_1=0}^{L-1} \sum_{j_2 \ne j_1 + \ell_1}  
m_2[j_1,j_1+\ell_2] m_1[j_2]
\label{iiii}
\end{align}
and
\begin{align}
& \frac{\gamma^3}{L^3} \sum_{j_2=0}^{L-1} \sum_{j_1 \ne j_2} 
m_1[j_1] m_1[j_2 + \ell_1] m_1[j_2 + \ell_2]
\label{jjjj}\\
& + \frac{\gamma^2}{L^2} \sum_{j_2=0}^{L-1} \sum_{j_1 \ne j_2}  
m_2[j_2+\ell_1,j_2+\ell_2] m_1[j_1].
\label{kkkk}
\end{align}

Finally, in Case \ref{case5} we have $i_1 - j_1$, $i_2 - j_2$, and $i_3 - j_3$ are all pairwise distinct. Consequently, the $x$ variables are always independent, and the expectation conditional on $s$ (letting $\Delta_{q} = i_q - j_q$, $q=1,2,3$),
\begin{align}
\sum_{j_1,j_2,j_3} s[\Delta_1] s[\Delta_2] s[\Delta_3] m_1[j_1] m_1[j_2] m_1[j_3];
\end{align}
since the $s[\Delta_q]$'s are pairwise independent, the expectation over $s$ then yields:
\begin{align} \label{llll}
\frac{\gamma^3}{L^3} \sum_{j_1,j_2,j_3} m_1[j_1] m_1[j_2] m_1[j_3].
\end{align}

Now we add all the terms from Cases \ref{case1} to \ref{case5}. Expressions \eqref{aaaa}, \eqref{ffff}, \eqref{hhhh}, \eqref{jjjj}, and \eqref{llll} sum to the expression:
\begin{align}
(\gamma a_x^1)^3.
\end{align}

Expressions \eqref{bbbb}, \eqref{cccc}, \eqref{dddd}, \eqref{gggg},\eqref{iiii}, and \eqref{kkkk} sum to the expression:
\begin{align}
\gamma a_x^1  \cdot 
( \gamma a_x^2[\ell_1] + \gamma a_x^2[\ell_2] + \gamma a_x^2[\ell_2-\ell_1]).
\end{align}
Finally, expression \eqref{eeee} is simply:
\begin{align}
\gamma a_x^3[\ell_1,\ell_2].
\end{align}

Now when $\sigma > 0$, we write $y = x^* + \ep$, and
\begin{align}
\E y[i_1] y[i_2] y[i_3] =\,\, & \E x^*[i_1] x^*[i_2] x^*[i_3] + \E x^*[i_1] \ep[i_2] \ep[i_3]
\nonumber \\
& + \E\ep[i_1] x^*[i_2] \ep[i_3] + \E \ep[i_1] \ep[i_2] x^*[i_3] 
+ \E \ep[i_1] \ep[i_2]\ep[i_3]
\nonumber \\
=\,\, & \E x^*[i_1] x^*[i_2] x^*[i_3] + \gamma  a_x^1 \cdot \sigma^2 \delta[i_2-i_3]
\nonumber \\
& + \gamma a_x^1 \cdot \sigma^2 \delta[i_3-i_1] 
+ \gamma a_x^1 \cdot \sigma^2 \delta[i_2-i_1] .
\end{align}

We conclude by averaging over all $i_1$, $i_2$, and $i_3$ with fixed values of $\ell_1 = i_2 - i_1$ and $\ell_2 = i_3 - i_1$.

\section{Proof of Proposition~\ref{prop:gamma}} \label{sec:proof_prop_gamma}
Refer to equations \eqref{eq:mean_micrograph}--\eqref{eq:ac3_micrograph} for expressions relating the moments of $y$ and those of $x$, and the parameters $\gamma$ and $\sigma$. First,  note that 
\begin{align*}
(a^1_y)^2 & = \frac{\gamma^2}{L^2}\sum_{i=0}^{L-1}\sum_{j=0}^{L-1}x[i]x[j].
\end{align*}
Similarly, using $a_x^2[-\ell] = a_x^2[\ell]$:
\begin{align*}
2\sum_{\ell = 1}^{L-1}a_y^2[\ell] & = \gamma \sum_{\ell = 1}^{L-1} \left(a_x^2[\ell] + a_x^2[-\ell]\right) \\
& = \frac{\gamma}{L}\sum_{i=0}^{L-1}\sum_{\ell = 1}^{L-1} x[i]\left(x[i+\ell] + x[i-\ell]\right) \\ &= \frac{\gamma}{L} \sum_{i = 0}^{L-1} \sum_{j = 0, j \neq i}^{L-1} x[i] x[j],
\end{align*}
where the last equality is obtained by noting that, given the summation bounds, the set of pairs $(i, i\pm\ell)$ and $(i, j)$ are the same over the valid range $\{0, \ldots, L-1\}^2$. To conclude, notice that $a_y^2[0]=\frac{\gamma}{L}\sum_{i=0}^{L-1}x[i]^2 + \sigma^2$ and combine.

\section{Proof of Proposition~\ref{prop:gamma_sigma}} \label{sec:proof_prop_gamma_sigma}

We prove that both $\sigma$ and $\gamma$ are identifiable from the observed first three moments of $y$. For convenience, we work with $\beta = \gamma / L$ rather than $\gamma$ itself. To this end, we construct two quadratic equations satisfied by $\beta$ and whose coefficients can be computed from observable quantities. Then, we show that these equations are independent, and hence that $\beta$ is uniquely defined. Given $\beta$, we can estimate $\sigma$ using Proposition~\ref{prop:gamma}.

Throughout the proof, it is important to distinguish between observed and unobserved values.
We denote the observed values by $E_i$ and $a_y^1,a_y^2,a_y^3$. We use $F_i$ to denote functions of the signal's autocorrelations (which are not directly observable).

Recall that $a_y^1 = \beta(\one^Tx)$ and $a_y^2[0] = \beta\|x\|^2+\sigma^2$, where $\one\in\RL$ is the vector of all-ones and $\|x\|=\sqrt{x[0]^2 + \cdots + x[L-1]^2}$ is the 2-norm. Consider the product $E_1$:
\begin{equation}\label{eq:E1}
\begin{split}
E_1 = a_y^1a_y^2[0] =  (\beta(\one^Tx))(\beta\|x\|^2+\sigma^2)  = \sigma^2a_y^1 + L\beta^2F_1,
\end{split}
\end{equation}
where $F_1 = a_x^3[0,0] + \sum_{j=1}^{L-1}(a_x^3[j,j] + a_x^3[0,j])$. 
The terms of $F_1$ can also be estimated from $a_y^3$, while taking the scaling and bias terms into account. This yields another observable, $E_2$:
\begin{align} 
E_2 & = a_y^3[0,0] + \sum_{j=1}^{L-1}(a_y^3[j,j] + a_y^3[0,j]) = L\beta F_1 + (2L+1)\sigma^2a_y^1. \label{eq:E2}
\end{align}
Therefore, from~\eqref{eq:E1} and~\eqref{eq:E2} we get:
\begin{equation} \label{eq:E12}
E_2\beta -(2L+1)\sigma^2\beta a_y^1 = E_1-\sigma^2a_y^1.
\end{equation}
Let $E_3 =a_y^2[0] + 2\sum_{\ell = 1}^{L-1}a_y^2[\ell]$; recall from Proposition~\ref{prop:gamma}:
\begin{equation} \label{eq:sigma2}
\sigma^2 = E_3 - (a^1_y)^2/\beta. 
\end{equation} 
Plugging into~\eqref{eq:E12} and rearranging, we get a first quadratic equation in $\beta$,
\begin{equation} \label{eq:quad1}
\mathcal{A}\beta^2 + \mathcal{B}\beta + \mathcal{C} = 0,
\end{equation}
where 
\begin{align*}
\mathcal{A} &= E_2 - (2L+1)a_y^1E_3, \\ 
\mathcal{B} &= -E_1 + (2L+1)(a_y^1)^3 + a_y^1E_3  , \\
\mathcal{C} &= -(a_y^1)^3.
\end{align*}
Importantly, these coefficients are observable quantities. As we assume throughout this proof that $x$ has nonzero mean, $a_y^1 \neq 0$ and we conclude that this equation is non-trivial.

Next, we derive the second quadratic equation for $\beta$. We notice that 
\begin{equation} \label{eq:E3}
E_4 = \frac{1}{L}(a_y^1)^3 = \frac{1}{L}\beta^3 (\one ^Tx)^3   = \beta^3 F_2,
\end{equation}
where $F_2 = \frac{1}{L}(\one ^Tx)^3$, and we can work out that:
\begin{equation*}
F_2 = a_x^3[0,0] + 3\sum_{j=1}^{L-1} \left(a_x^3[j,j] + a_x^3[0,j]\right) + 6\sum_{1\leq i < j\leq L-1}a_x^3[i,j].
\end{equation*}
Once again, $F_2$ can be estimated from $a_y^3$, taking bias and scaling into account:
\begin{align}
E_5 & = a_y^3[0,0] + 3\sum_{j=1}^{L-1} \left(a_y^3[j,j] + a_y^3[0,j]\right) + 6\sum_{1\leq i < j\leq L-1}a_y^3[i,j]  = L \beta F_2 + (6L-3)\sigma^2a_y^1.
\end{align}
Consider the following ratio:
\begin{equation*} 
\frac{E_5}{E_4} = \frac{L}{\beta^2} + \frac{(6L-3)\sigma^2a_y^1}{E_4}.
\end{equation*}
From the latter, we deduce:
\begin{equation*}
\sigma^2 = \frac{E_5}{a_y^1(6L-3)}  - \frac{LE_4}{\beta^2a_y^1(6L-3)}.
\end{equation*}
Using~\eqref{eq:sigma2} and rearranging, we get the second quadratic:
\begin{equation} \label{eq:quad2}
\mathcal{D}\beta^2 + \mathcal{E}\beta + \mathcal{F} = 0,
\end{equation}
where
\begin{align*}
\mathcal{D} &= E_3 - \frac{E_5}{a_y^1(6L-3)}, \\ 
\mathcal{E} &= -(a_y^1)^2, \\
\mathcal{F} &= \frac{LE_4}{a_y^1(6L-3)}.
\end{align*}
It is also non-trivial since $E_4 \neq 0$.

To complete the proof, we need to show that the two quadratic equations~\eqref{eq:quad1} and~\eqref{eq:quad2} are independent. To this end, it is enough to show that the ratios between coefficients differ. 
From~\eqref{eq:quad1} and~\eqref{eq:E1}, we have:
\begin{equation*}
\begin{split}
\frac{\mathcal{B}}{\mathcal{C}} = \frac{E_1 - (2L+1)(a_y^1)^3 - a_y^1E_3}{(a_y^1)^3} = \frac{a_y^2[0] - (2L+1)(a_y^1)^2 - E_3}{(a_y^1)^2}.
\end{split}
\end{equation*}
In addition, using~\eqref{eq:E3},
\begin{equation*}
\frac{\mathcal{E}}{\mathcal{F}} = \frac{(3-6L)(a_y^1)^3}{LE_4} = 3 - 6L . 
\end{equation*}
For contradiction, suppose that the quadratics are dependent. Then, $\frac{\mathcal{B}}{\mathcal{C}} =\frac{\mathcal{E}}{\mathcal{F}} $, that is, 	
\begin{equation*}
a_y^2[0] - (2L+1)(a_y^1)^2 - E_3 = (a_y^1)^2(3-6L).
\end{equation*}
Rewriting the identity in terms of $x$ and dividing by $\beta$ we get:
\begin{equation} \label{eq:cond}
4(L-1)\beta (\1^\top x)^2  - (\1^\top x)^2 + \|x\|^2 = 0.
\end{equation}	
For generic $x$,  this polynomial equation is not satisfied so that the quadratic equations are independent. 
Furthermore, from the inequality $L\|x\|^2 \ge (\1^\top x)^2$ it follows immediately that the equations must be independent so long as
\begin{equation*}
\beta > \frac{1}{4L}.
\end{equation*}

\section{Proof of Proposition~\ref{prop:uniqueness_poisson}} \label{sec:proof_uniqueness_poisson}

We first note that $\gamma a_x^1$, $\gamma a_x^2[\ell] + \sigma^2 \delta[\ell]$, and $\gamma a_x^3[\ell_1,\ell_2]$ can be computed directly form the observed autocorrelations~\eqref{eq:mean_micrograph2},~\eqref{eq:ac2_micrograph2} and~\eqref{eq:ac3_micrograph2}.
Indeed, recovering  $\gamma a_x^1$ and $\gamma a_x^2[\ell] + \sigma^2 \ell[\ell]$ is immediate from \eqref{eq:mean_micrograph2} and \eqref{eq:ac2_micrograph2}, and $\gamma a_x^3[\ell_1,\ell_2]$ then follows from
\begin{align}
\gamma a_x^3[\ell_1,\ell_2]
= a_y^3[\ell_1,\ell_2] - (a_y^1)^3
- a_y^1 \cdot \big(a_y^2[\ell_1]  
+ a_y^2[\ell_2] + a_y^2[\ell_1-\ell_2]- 3(a_y^1)^2 \big).
\end{align}

Let us assume that $x$ is generic. Indeed, we observe the product:
\begin{align}
L^2(\gamma a_x^1) (\gamma a_x^2[1])
&= \gamma^2\left( \sum_{i} x[i] \right) \left(\sum_{j} x[j]x[j+1] \right)
\nonumber \\
&= \gamma^2\sum_{j} \sum_{i} x[i]x[j]x[j+1]
\nonumber \\
&=  \gamma^2\sum_{j} \sum_{\ell} x[j+\ell]x[j]x[j+1]
\nonumber \\
&=  \gamma^2\sum_{\ell \ge 0} \sum_{j} x[j+\ell]x[j]x[j+1] 
+ \gamma^2\sum_{\ell < 0} \sum_{j} x[j+\ell]x[j]x[j+1]
\nonumber \\
&=  \gamma^2\sum_{\ell=0}^{L-1} a_x^3[1,\ell] 
+ \gamma^2\sum_{\ell > 0} \sum_{j} x[j-\ell]x[j]x[j+1]
\nonumber \\
&=  \gamma^2\sum_{\ell=0}^{L-1} a_x^3[1,\ell] 
+ \gamma^2\sum_{\ell > 0} \sum_{j} x[j]x[j+\ell]x[j+\ell+1]
\nonumber \\
&= \gamma\left( \sum_{\ell=0}^{L-1} \gamma a_x^3[1,\ell]
+ \sum_{\ell=1}^{L-2} \gamma a_x^3[\ell,\ell+1]  \right).
\end{align}
Since we also observe $\sum_{\ell=0}^{L-1} \gamma a_x^3[1,\ell] + \sum_{\ell=1}^{L-2} \gamma a_x^3[\ell,\ell+1]$, we can form the ratio and solve for $\gamma$.

\begin{align}
\gamma = L \frac{(\gamma a_x^1) (\gamma a_x^2[1])}
{\sum_{\ell=0}^{L-1} \gamma a_x^3[1,\ell] 
	+ \sum_{\ell=2}^{L-1} \gamma a_x^3[\ell,\ell+1]},
\end{align}
Then, similarly to Appendix~\ref{sec:proof_prop_gamma}, one can solve for $\sigma^2$:
\begin{align}
\sigma^2 = a_y^2[0] + 2\sum_{\ell = 1}^{L-1}a_y^2[\ell]-\frac{L (a^1_y)^2}{\gamma}
- (2 L - 1) (a_y^1)^2.
\end{align}
Once $\gamma$ and $\sigma$ were computed, one can recover $x$ from $a_x^2$ and $a_x^3$  by Proposition~\ref{prop:uniqueness}.

\end{document}